\documentclass[a4paper,11pt]{article}

\usepackage{jheppub}
\usepackage{lineno}

\usepackage[utf8]{inputenc}
\usepackage{lipsum}
\usepackage{xcolor}
\usepackage{amsthm,amsfonts,mathrsfs}
\usepackage{tabularx}
\usepackage{booktabs}
\usepackage{subfigure}
\usepackage{tikz}
\usepackage[normalem]{ulem}

\usepackage{latexsym}
\usepackage{ulem}

\usepackage{mathtools} 
\usepackage{dsfont}
\usepackage{slashed}
\usepackage[thinc]{esdiff}
\usepackage{comment}
\usepackage{cancel}
\usepackage{xcolor}
\usepackage{tikz-feynman}
\usepackage{gensymb}
\usepackage{multirow}
\usepackage[shortlabels]{enumitem}

\usepackage{cases} 
\usepackage{soul} 

\usepackage{amssymb}

\newcommand{\be}{\begin{eqnarray}}
\newcommand{\ee}{\end{eqnarray}}

\newcommand\De{\Delta}

\newcommand{\bea}{\begin{eqnarray}}
\newcommand{\eea}{\end{eqnarray}}

\begin{document}

\title{Ultraviolet completion of the inflationary paradigm}

\author{Leonardo Modesto,}
\emailAdd{leonardo.modesto@unica.it}

\author{Lorenzo Orlando}
\emailAdd{lorenzo.orlando@ca.infn.it}

\affiliation{Dipartimento di Fisica, Universit\`a di Cagliari, Cittadella Universitaria, 09042 Monserrato, Italy}
\affiliation{I.N.F.N, Sezione di Cagliari, Cittadella Universitaria, 09042 Monserrato, Italy}

\abstract{After an exhaustive introduction highlighting the strengths and weaknesses
of the non-local models proposed so far as ultraviolet completions of the Starobinsky theory, we propose a new nonlocal completion of a general $f(R)$ theory (in the Einstein's frame) suitable for driving inflation in the early universe consistently with observations.
The nonlocal theory shares with $f(R)$ the same background solutions and the same equations of motion for perturbations at linear and nonlinear level. Therefore, the classical cosmological observables are not affected by the nonlocal operators needed for the quantum completion. Our construction applies to any local action written in the Einstein's frame, but we will provide the details only for two explicit examples: the Starobinsky model and a general $f(R)$ theory.
The new model overcomes the incompatibility of (super-)renormalizability and stability present in the previous proposals based on the explicitly known asymptotically polynomial form factors. 

Since the nonlocal theory is at least super-renormalizable, but can also be finite depending on the details of the model, this work shows the consistency of the inflationary paradigm with a well-defined quantum theory of gravity at high energy. We could rephrase the latter statement saying that the success of the $f(R)$ theories in cosmology can be traced back to the existence of an ultraviolet completion that preserves all the classical features. 
The inflationary paradigm survives, or it is actually insensitive to quantum gravity, because it is an exact solution of quantum gravity, up to perturbative corrections. }

\maketitle

\section{Introduction}

There are three main issues in the standard cosmological model: the horizon problem, the flatness problem, and the singularity problem. Furthermore, the spectrum of the cosmic microwave background radiation (CMB) is 
quasi-scale-invariant: a feature not easy to explain in Einstein's gravity coupled to the standard model of particles physics 
(in such respect, one viable scenario is based on the Higgs boson 
\cite{PhysRevD.40.1753, Bezrukov:2007ep,Barvinsky:2008ia,Garcia-Bellido:2008ycs,DeSimone:2008ei,Barbon:2009ya,Bezrukov:2010jz,Bezrukov:2011gp,He:2018gyf}. 
Therefore, in order to avoid a purely quantum gravity solution of all these issues\footnote{For a quantum gravity alternative to inflation see for example \cite{Modesto:2022asj, Calcagni:2022tuz}.}, physicists proposed the inflationary scenario that, in our days, we can more properly define: inflationary paradigm \cite{Linde:1981mu, Guth:1980zm, Starobinsky:1979ty, Biswas:2012bp, Cai:2015emx, Koshelev:2016xqb, Nojiri:2017ncd, Nojiri:2019dio, Bombacigno:2024lud, Nojiri:2026hij}. Indeed, the features of such model are quite general and shared by a very large class of theories among which the most successful one was proposed by Starobinsky in $1979$ \cite{Starobinsky:1979ty, Starobinsky:1980te, Starobinsky:1983zz}. 

In this paper, we do not propose a further model or criticize the inflationary framework, but we try to understand the observational success of the paradigm despite the energy scale of inflation being not so much below the quantum gravity scale: $10^{15}$ GeV versus $10^{19}$ GeV.
In other words, we would like to understand whether it is really possible to uplift a general $f(R)$ model to a consistent quantum gravity theory while preserving all the classical results of the former.
The outcome would be a recipe suitable to be applied at our favorite theory for the early Universe in order to obtain its ultraviolet completion, but without losing all the perturbative results of the $f(R)$ model. In particular the CMB prediction of the specific model. 
We will mainly focus on a non-local extension of the Starobinsky minimal $f(R)$ model, but the power of the result lies precisely in its capability to stabilize inflation from any interference from quantum gravity. 
However, It is straightforward to extend the proposal to higher derivative theories too \cite{Asorey:1996hz, Modesto:2015ozb, Anselmi:2017yux, Anselmi:2017lia, Anselmi:2018kgz, Anselmi:2020lpp}.

Regarding nonlocal gravity, N.V. Krasnikov in $1987$ \cite{Krasnikov:1987yj}, in order to overcome the ghost issue of Stelle's quadratic gravity \cite{Stelle:1976gc}, proposed a new class of gravitational theories tree-level unitary and higher derivative at the same time. 
Two years later, Yu.V. Kuz'min modified the theory to a one well-defined in the high energy regime, namely a theory consistent with the power counting renormalizability \cite{Kuzmin}. 

About thirty years later, a class of weakly nonlocal gravitational theories \cite{modesto} was generalized to any dimension and proved to be finite, first in odd dimension \cite{Modesto:2014lga} and afterwards in even dimension \cite{Modesto:2014lga, Koshelev:2025pxg}.
Unlike the higher derivative theories \cite{JulveTonin, Stelle:1976gc}, weakly nonlocal gravitational theories are naturally ghost free without the need of special prescriptions. Indeed, the usual Feynman $i \epsilon$ prescription guarantees the Cutkosky rules together with perturbative Unitarity in Minkowski spacetime \cite{Briscese:2018oyx, Koshelev:2021orf, Briscese:2021mob}. Moreover, all the solutions of Einstein's gravity and their stability properties are preserved in nonlocal classical gravity at any perturbative order \cite{StabilityMinkAO, StabilityRicciAO}. 
Unlike the higher derivative theories \cite{JulveTonin, Stelle:1976gc}, weakly nonlocal gravitational theories are naturally ghost free without the need of special prescriptions. Indeed, the usual Feynman $i \epsilon$ prescription guarantees the Cutkosky rules together with perturbative Unitarity in Minkowski spacetime \cite{Briscese:2018oyx, Koshelev:2021orf, Briscese:2021mob}. Moreover, all the solutions of Einstein's gravity and their stability properties are preserved in nonlocal classical gravity at any perturbative order \cite{StabilityMinkAO, StabilityRicciAO}. 
Unlike the higher derivative theories \cite{JulveTonin, Stelle:1976gc}, weakly nonlocal gravitational theories are naturally ghost free without the need of special prescriptions. Indeed, the usual Feynman $i \epsilon$ prescription guarantees the Cutkosky rules together with perturbative Unitarity in Minkowski spacetime \cite{Briscese:2018oyx, Koshelev:2021orf, Briscese:2021mob}. Moreover, all the solutions of Einstein's gravity and their stability properties are preserved in nonlocal classical gravity at any perturbative order \cite{StabilityMinkAO, StabilityRicciAO}. 
Unlike the higher derivative theories \cite{JulveTonin, Stelle:1976gc}, weakly nonlocal gravitational theories are naturally ghost free without the need of special prescriptions. Indeed, the usual Feynman $i \epsilon$ prescription guarantees the Cutkosky rules together with perturbative Unitarity in Minkowski spacetime \cite{Briscese:2018oyx, Koshelev:2021orf, Briscese:2021mob}. Moreover, all the solutions of Einstein's gravity and their stability properties are preserved in nonlocal classical gravity at any perturbative order \cite{StabilityMinkAO, StabilityRicciAO}. 
Unlike the higher derivative theories \cite{JulveTonin, Stelle:1976gc}, weakly nonlocal gravitational theories are naturally ghost free without the need of special prescriptions. Indeed, the usual Feynman $i \epsilon$ prescription guarantees the Cutkosky rules together with perturbative Unitarity in Minkowski spacetime \cite{Briscese:2018oyx, Koshelev:2021orf, Briscese:2021mob}. Moreover, all the solutions of Einstein's gravity and their stability properties are preserved in nonlocal classical gravity at any perturbative order \cite{StabilityMinkAO, StabilityRicciAO}. 
Unlike the higher derivative theories \cite{JulveTonin, Stelle:1976gc}, weakly nonlocal gravitational theories are naturally ghost free without the need of special prescriptions. Indeed, the usual Feynman $i \epsilon$ prescription guarantees the Cutkosky rules together with perturbative Unitarity in Minkowski spacetime \cite{Briscese:2018oyx, Koshelev:2021orf, Carone:2016eyp, Briscese:2021mob, Pius:2016jsl}. Moreover, all the solutions of Einstein's gravity and their stability properties are preserved in nonlocal classical gravity at any perturbative order \cite{StabilityMinkAO, StabilityRicciAO}. 
Macro-causality is also under control \cite{causality}, while we expect harmless violations of micro-causality at the non-locality scale \cite{brisceseCausality}. 
Regarding the application to the early Universe cosmology, other authors have done an excellent job in studying the theoretical and observational features of nonlocal gravitational theories
\cite{Biswas:2016etb, Biswas:2012bp, calcagni, Koshelev:2016xqb, SravanKumar:2018dlo}. 

The main efforts concerned the selection of nonlocal form factors able to ensure stability of scalar and tensor perturbations around a maximally symmetric spacetime (MSS). Using the technical machinery developed in 
\cite{Biswas:2016etb, Biswas:2016egy}, a form factor satisfying the required properties and compatible with a MSS during the inflationary stage of the Universe was introduced in \cite{Koshelev:2016xqb}. Afterwards, slight improvements were implemented in \cite{SravanKumar:2018dlo}, leading to form factors \eqref{form_factor_S}-\eqref{form_factor_C} more suitable around a quasi-dS background. The latter took into account the contribution (small during inflation, $R_{\text{infl}}\gg 6M^2$) of the scalaron into the tensor form factor, and in considering the not negligible contribution quadratic in the Weyl curvature for the scalar form factor. Indeed, the scalar form factor computed in \cite{Koshelev:2016xqb} was a good proposal only around the Minkowski background.

The main issue of the models proposed till now regards the stability at any perturbative order. 
In particular, in order for the theory to be ghost-free around any background, we need to consider those form factors to be always valid \cite{Koshelev:2022olc}. However, as it was suggested in \cite{calcagni} and as it will be shown here, this first way breaks good renormalizability properties of this class of non-local theories.
Alternatively, if we want to preserve renormalizability, one has to deal with new masses in the spectrum: an infinite number of complex conjugate masses \cite{Biswas:2016etb, Koshelev:2020fok}.
As shown in this paper, it is crucial that in nonlocal theories, contrary to what happens in the higher derivative ones, the complex instabilities are infinite thus making the background solution decay instantaneously.

The current paper addresses the following issues. In Section \ref{local_infl}, we expand about the  
theoretical problems of the Starobinsky original theory \cite{Starobinsky:1979ty, Starobinsky:1980te, Starobinsky:1983zz}: the renormalizability issue, but most importantly its inconsistency as an effective field theory. 

In Section \ref{nonlocal_theories}, we show that the classes of Starobinsky nonlocal theories compatible with the Starobinsky inflation proposed so far are either: non-renormalizable (because of an extra curvature operator needed to have stable perturbations around any background) or classically unstable. 
In order to solve the problems pointed out, in Section \ref{NLGM_approach} we propose a completion of the Starobinsky theory based on a nonlocal unified theory of gravity and matter \cite{Modesto:2021ief} consistent with super-renormalizability or finiteness \cite{Calcagni:2023goc} and unitarity. In particular, the stability properties of the nonlocal theory are the same of the Starobinsky model. Therefore, the nonlocal theory provides exactly the same predictions of the Starobinsky theory.
The above construction can be extended to any $f(R)$ theory in the Einstein's frame. Therefore, all inflationary models based on a general classical $f(R)$ theory are safe from relevant quantum gravity corrections. 

Finally, in Section \ref{renorm_proof}--\ref{finite_number} we prove the renormalizability of the aforementioned class of nonlocal theories \eqref{NLGM_fR} and we draw our conclusions.

Along the paper, we will assume the metric signature to be mostly plus: $(-+++)$. We will work in natural units and use the standard definitions for the Planck mass $m_{p}$ and the reduced Planck mass $M_{\rm p}$, i.e. 
\be
    m_{\rm p}=\sqrt{\frac{\hbar c}{G}} \, , \quad 
    M_{\rm p}=\frac{m_{\rm p}}{\sqrt{8\pi}}  \, , \quad \kappa^2 = 16 \pi G = \frac{16 \pi}{m_{\rm p}^2}  = \frac{2}{M_{\rm p}^2}  \, .
\ee

\section{The Starobinsky theory: pros and cons}\label{local_infl} 
The higher-derivative model we are going to recap and analyze in this section was proposed by A. Starobinsky in \cite{Starobinsky:1979ty, Starobinsky:1980te, Starobinsky:1983zz} in 1979, and the action reads:
\begin{equation}
\label{localR^2}
    S=\int d^{4}x\sqrt{-g}\left( \frac{M_{\rm p}^{2}}{2}R+\frac{M_{\rm p}^{2}}{12M^{2}}R^{2}\right),
\end{equation}
where $M_{\rm p}$ is the reduced Planck mass and $M$ is the mass of an additional degree of freedom usually called scalaron or curvaton. Indeed, as evident from the (gauge-independent part of the) propagator for the quantum fluctuation $h_{\mu\nu}=\kappa^{-1}(g_{\mu\nu}-\bar g_{\mu\nu})$, 
\begin{equation}
\label{appendix_propagator}
    \mathcal{O}^{-1}(k)=-\left[\frac{P^{(2)}}{k^2}-\frac{P^{(0)}}{(D-2)k^2\left(1+\frac{k^{2}}{M^{2}}\right)}\right] , 
\end{equation}
the spin zero contribution has two poles: the massless one, which together with the spin two projector gives the right number of components for the massless graviton, and another massive pole due to the presence of a scalar particle in the spectrum. It is the extended spectrum of the Starobinsky theory in comparison to Einstein's gravity to make possible an inflationary era driven by the scalaron.  
When $R^2$ dominates over $R$, inflation takes place and the metric is quasi deSitter. 
The Starobinsky theory, with the addition of 1-loop quantum gravitational corrections, provides an interpolation between the very early Universe and the radiation-dominated stage, while a graceful exit into the matter-dominated stage is ensured by the scalaron. 
From a phenomenological point of view, we recall that Starobinsky theory produces predictions in agreement with the most recent observational bounds \cite{Planck:2018jri}, as it was shown in \cite{DeFelice:2010aj} for the scalar spectral index $n_{s}$ and for the tensor-to-scalar ratio $r$.

For future purposes, it is useful to recall that any $f(R)$ theory, 
\begin{equation}\label{fR}
    S= \frac{M_{\rm p}^2}{2} \int d^{4}x\sqrt{-g} f(R),
\end{equation}
can be equivalently written introducing an on-shell auxiliary scalar field $\varphi=R$, namely the theory is written in the so-called Jordan frame as follows, 
\be
\label{Jordan_frame}
S_{\text{JF}} = \dfrac{M_{\rm p}^2}{2} \int d^{4}x\sqrt{-g} \left[F(\varphi)R-U(\varphi)\right] \, , \quad 
    F(\varphi)=\dfrac{df(\varphi)}{d\varphi},\quad U(\varphi)=F(\varphi)\varphi-f,
\ee
which shows a non-minimal coupling between gravity and a real scalar field. Hence, by a conformal transformation we can turn \eqref{fR} into a minimally-coupled scalar-tensor theory (Einstein frame). The result reads:
\begin{eqnarray}
\label{Einstein_frame}
    S_{\text{EF}}=\int d^4 x\sqrt{-\tilde{g}}\left[\frac{M_p^2}{2}\tilde R-\frac{1}{2}\tilde g_{\mu\nu}\partial^{\mu}\phi\partial^{\nu}\phi-V(\phi)\right], \quad \phi=\sqrt{\frac{3}{2}}\,M_{\mathrm{p}}\ln F(\varphi).
\end{eqnarray}
\subsection*{The renormalizability and effective theory issues} 
Although the $R+R^2$ theory agrees with the observations, from a theoretical point of view it has two important drawbacks: it is neither a renormalizable field theory nor an effective field theory.
\emph{Renormalizability} is the main guiding principle in quantum field theory that allows to restrict the class of field theories into the ones predictive at each perturbative order. Indeed, in renormalizable theories only a finite number of coupling constants are affected by an infinite renormalization. Therefore, only a finite number of constants need to be measured in order to have the full quantum action starting from a classical one. 
Alternatively, we deal with non-renormalizable theories, which are consistent as long as the higher derivative operators are suppressed by a specific mass scale. This is the effective theory approach. However, the validity of these theories is very subtle because all the low energy operators have to be chosen with meticulous precision in order to avoid any ghost state at whatsoever scale \cite{Simon:1990ic,  deRham:2010kj}. Indeed, the truncation of a theory having a ghost state at a very high energy, in comparison with the experimental scale, suffers of an instantaneous vacuum instability even if the two scales are many orders of magnitude apart.
Coming to the point, the theory \eqref{localR^2} is non-renormalizable because already at $1$-loop we must include counterterms, absent in the classical Lagrangian, with (up to) $2 N_V+4$ derivatives, i.e.  
\begin{equation}
    \frac{1}{\varepsilon}\int d^{4}x\sqrt{-g}\,\mathcal{O}_V,\quad \mathcal{O}_V\sim\partial^{2N_V+4},
   \label{CTS0}
\end{equation}
where $N_V$ stays for the number of vertices of a given diagram while the schematic notation for the operators $\mathcal{O}_V$ highlights how many derivatives they contain at most.
As an explicit example, counterterms up to the form (or the analogous ones with less derivatives)
\begin{equation} 
 \{ \mathcal R^{N_V+2},R^{N_V}\nabla R,\cdots,\mathcal{R}\nabla^{2N_V}\mathcal{R} \}
   \label{CTS}
\end{equation}
and their corresponding made of any possible permutation between covariant derivatives and curvatures
are to be expected, provided that the diagrams taken into account have a compatible number of external gravitons. 
In (\ref{CTS}), $\mathcal{R}$ generically stays for $\{R,\,R_{\mu\nu},\,R_{\mu\nu\alpha\beta}\}$.
The reason for the proliferation of so many counterterms can easily be traced back to the structure of the propagator and the vertices. In fact, the spin two component of the propagator scales like $k^{-2}$, while some vertices like $k^4$. Therefore, at one-loop the increase of vertices makes the amplitude more and more divergent (see Section \ref{Toy_model} for more details).

In light of the non-renormalizability, one could think the Starobinsky theory as an effective theory.
According to the \emph{naturalness argument}, all the coupling constants in an effective theory are chosen to be of order one in units of the cut-off mass scale.
Therefore, the theory shows only one cut-off mass scale, while all the couplings are dimensionless coefficients. 
 As we are going to show, it turns out that the Starobinsky theory can not be an effective field theory \cite{Weinberg:1978kz}.
 In other words, it is not possible to find an energy regime such that all the operators with more than four derivatives are suppressed in comparison to $R^2$ and the action simplifies to:
 \begin{equation}\label{effect1}
       S=\int d^4 x\sqrt{-g}\left[\frac{M_{p}^{2}}{2}R+\epsilon_{2}R^{2}\right],\quad \epsilon_2\sim1 \, .
   \end{equation}
Indeed, let us assume \eqref{effect1} to be a truncation at the order $R^2$ of the 
following general action,  
\begin{equation}\label{effect2}
       S=\int d^4 x\sqrt{-g}\left[\frac{M_{p}^{2}}{2}R+\sum_{n=2}^{\infty}
     \frac{  \epsilon_{n} }{M_{\rm p}^{2 n  - 4} }
       \sum_{\sigma}\mathcal{R}^{n_{\sigma}}\right].
   \end{equation}
where the sum is on all possible ways of combining $n_\sigma$ covariant tensors $\mathcal{R}$, 
and $\epsilon_{n},\,n\geq 2$ are coefficients satisfying the naturalness identification, namely $\epsilon_{n}\sim 1$.
\begin{equation}
\label{naturalness}
    \epsilon_{n}\sim 1  
    \, . 
\end{equation}
In order to show that there is no energy regime at which \eqref{effect1} is an effective description of \eqref{effect2}, we look for an energy scale where the $R^{2}$ term dominates. In other words, $R^2$ has to dominate over the Einstein Hilbert action and the next term $\epsilon_{3}R^{3}$ as well as on any other subsequent operator. 
In terms of inequalities, we can rephrase the last statement as:
\begin{equation}\label{inequalities}
    \begin{cases}
        \epsilon_{2} {\mathcal R}^{2}\gg M_{\rm p}^{2} {\mathcal R}\\ \\
        \epsilon_{2} {\mathcal R}^{2}\gg \dfrac{\epsilon_{3}}{M^2_{\rm p}} {\mathcal R}^{3}
    \end{cases}
    \quad
    \stackrel{\eqref{naturalness}}{\implies}
    \quad
    \begin{cases}
        {\mathcal R} \gg M_{\rm p}^{2}  \\
        {\mathcal R}^{2}\gg \dfrac{{\mathcal R}^{3}}{M_{\rm p}^{2}} 
    \end{cases}
    \quad
\stackrel{E\sim {\mathcal R}^{1/2}}{\implies} 
  \quad
   \begin{cases}
       E\gg M_{\rm p} \\
       E\ll M_{\rm p}  
    \end{cases}
\end{equation}
where we identified the energy scale squared with the curvature ${\mathcal R}$, namely $E^2\sim \mathcal{R}$. The last condition in (\ref{inequalities}) implies that we cannot have a Starobinsky regime in the effective field theory approach.
Conversely, ignoring the definition of effective theory based on (\ref{naturalness}), thus, adding at least  
another mass scale in the coefficient $\epsilon_2$, namely:
\be
\epsilon_{2}=\dfrac{M_{\rm p}^{2}}{M^2} \quad {\rm and} \quad  \epsilon_n \sim 1 \quad {\rm for} \quad n>2,
\ee

the inequalities (\ref{inequalities}) turn into: 
\begin{equation}
 \begin{cases}
        \epsilon_{2} {\mathcal R}^{2}\gg M_{\rm p}^{2} {\mathcal R}\\ \\
        \epsilon_{2}R^{2}\gg
         \dfrac{  \epsilon_{n} }{M_{\rm p}^{2 n  - 4} }
        \mathcal{R}^{n_\sigma} 
        \quad (n > 2) 
    \end{cases}
\quad    
\stackrel{E\sim {\mathcal R}^{1/2}}{\Longrightarrow}  
\quad 
\begin{cases}
         E\gg M\\ \\
       \dfrac{E}{M_{\rm p}}\ll \left(\dfrac{M_{\rm p}^2}{M^2}\right)^{\frac{1}{2n-4}}.
\end{cases}
\end{equation}
Setting $n=3$, we get \footnote{The value $M \approx 1.3 \cdot 10^{-5}M_{\rm p}$ is in agreement with the observational data \cite{Koshelev:2016xqb}.} $M\ll M_{\rm p}$. 
Finally, we notice that the $R^{2}$ term dominates with respect to any higher derivative term. Indeed, 
\begin{equation}
   \frac{E}{M_{\rm p}}\ll \left(\dfrac{M_{\rm p}^2}{M^2}\right)^{\frac{1}{2n-4}} \quad 
   \overset{ n\rightarrow \infty }{\longrightarrow} \quad 1 \quad \mbox{from above} .
\end{equation}
Hence, the energy regime where inflation is described by \eqref{localR^2} is:
\begin{equation}
   \boxed{ M\ll E\ll M_{\rm p}} \, . 
\end{equation}
So far, we proved the following statement: \emph{only a theory with a second scale $M$ different from the cut-off one, as in the Starobinsky case, can allow for inflation}. 
In other words, according to the common definition of effective theory, the Starobinsky model cannot be seen as such.  
 Thus, the only way to make the Starobinsky model consistent with small-scale physics is to look for its ultraviolet completion. Since (\ref{localR^2}) is already incomplete from the geometric point of view, the natural and minimal completion of (\ref{localR^2}) is provided by the Stelle action \cite{Stelle:1976gc}, 
\begin{equation}\label{Stelle}
    S=\int d^{4}x\sqrt{-g}\left(\frac{M_{\rm p}^{2}}{2}R+\frac{M_{\rm p}^{2}}{12M^2}R^{2}+c_{2}R_{\mu\nu}R^{\mu\nu}\right) \, , 
\end{equation}
where the addition of the Ricci squared tensor $\mathbf{Ric}^2$ (or equivalently the Weyl squared tensor in a different base) is sufficient in order to make the Starobinsky action renormalizable \cite{Stelle:1976gc}. 
The spectrum of the theory shows a spin two massive ghost that implies an instantaneous decay of the Minkowski vacuum in gravitons of positive energy and ghosts with negative energy. Therefore, the theory seems unstable at classical as well as at quantum level. This issue is very severe because even in an effective theory the spectrum should be ghost free at any energy scale.
Assuming that the mass of the ghost is much greater than the measurable energy scale is in itself useless because, as just said, the vacuum will immediately decay regardless of the difference in mass. 
However, the theory is unitary whether a proper prescription is introduced  \cite{Anselmi:2017yux, Anselmi:2017lia}. In short, it consists in removing by hand the ghost from the asymptotic spectrum of the theory, while a new prescription (alternative to the Feynman prescription) is needed in evaluating the loop diagrams in order to avoid the contribution of the ghost to the Cutkosky rules. 
Therefore, the imaginary part of the amplitude is zero and unitarity secured. 

So far so good, but in this paper our aim is to abandon the locality principle and further consider and study weakly nonlocal extensions of the Starobinsky theory. These theories, unlike the Stelle's or higher derivative theories, do not exhibit any instability and unitarity is guaranteed at the quantum level in Minkowski spacetime. On the other hand, the asymptotically polynomial character in the ultraviolet limit ensures the convergence of amplitudes with a number of loops greater than or equal to two. One-loop amplitudes can be made finite with a simple extension of the theory.

\section{A non-local completion of the Starobinsky theory}\label{nonlocal_theories}
From this section onwards, we will focus on nonlocal theories. We will review previously studied theories \cite{Koshelev:2016xqb, SravanKumar:2018dlo}, highlighting some of their shortcomings, and finally propose a new theory consistent with the inflationary paradigm at any perturbative order.
A non-local theory suitable for Cosmology and extensively studied in the past is\footnote{The theory written in Weyl basis simplifies the search for exact solutions because the Weyl tensor vanishes on any FLRW background.} 
\begin{equation}\label{NLtheory}
    S=\frac{M_{\rm p}^{2}}{2}\int d^{4}x\sqrt{-g}\left[R+R\gamma_{\rm S}(\square)R+C_{\mu\nu\rho\sigma}\gamma_{\rm C}(\square)C^{\mu\nu\rho\sigma}+\mathcal{V}(C)\right],
\end{equation}
which is unitary at any perturbative order in the loop expansion \cite{Briscese:2018oyx, Briscese:2021mob}
and super-renormalizable or finite at quantum level if we introduce the following potential, 
\begin{equation}
    \mathcal{V}(C)= s_1 C^2\Box^{n-4} C^2+s_2 C_{\mu\nu\rho\sigma}C^{\alpha\beta\gamma\delta}\Box^{n-4} C_{\alpha\beta\gamma\delta}C^{\mu\nu\rho\sigma} .
\end{equation}
The one-loop finiteness is achieved for a particular choice of front coefficients $s_1$ and $s_2$ \cite{Modesto:2014lga}. The exponent $n$ is a positive integer number properly selected in order to achieve super-renormalizability \cite{Kuzmin, modesto}. 
As originally proposed in \cite{Krasnikov:1987yj}, the absence of ghosts in Minkowski spacetime forces 
$\gamma_{\rm S}(\Box),\,\gamma_{\rm C}(\Box)$ to be entire functions, while the power counting renormalizability requires the form factors to be asymptotically polynomial. 
A particular choice of the form factor with the required properties is:
\begin{align}
 \gamma_{\rm S}(\Box)&=-\dfrac{1}{6\Box}\left[e^{H_0(\Box)}\left(1-\dfrac{\Box}{M^2}\right)-1\right], & H_0(\Box)&=\dfrac{\alpha}{2}\{\ln p^{2}(\square)+\Gamma[0,\,p^{2}(\square)]+\gamma_{E}\},
   \nonumber  \\
   \gamma_{\rm C}(\Box)& = \dfrac{e^{H_2(\Box)}-1}{2\Box}, & H_2(\Box)&=\dfrac{\alpha}{2}\{\ln q^{2}(\square)+\Gamma[0,\,q^{2}(\square)]+\gamma_{E}\} \, ,
 \label{general_form}
\end{align}
  where, for the sake of simplicity, we have denoted the argument of the entire functions $H_0, H_2$ by:
    \begin{equation}\label{adim_Box}
        \frac{\Box}{\Lambda_{*}} \,\, \rightarrow \,\, \Box \, , 
    \end{equation}
while 
  $p(z)$ and $q(z)$ are respectively polynomials of degree $n-2,\,n-1$.
     Finally, $\Lambda_{*}$ is the fundamental \emph{non-locality mass scale} that satisfies 
    \begin{equation}
       M\ll \Lambda_{*}\lesssim M_{\rm p} 
    \end{equation}
   in accordance to early Cosmology data \cite{Koshelev:2016xqb}. 

However, as we are going to show, it is not possible to satisfy the above requirements of super-renormalizability and stability together with the Starobinsky inflation. 

Let us start with the stability issue. 
In previous works \cite{Koshelev:2016xqb, SravanKumar:2018dlo}, the form factors $\gamma_S,\,\gamma_C$ were fixed in order to have the inflationary Starobinsky exact solution as well as the linear equations of motion for the perturbations around the quasi-de Sitter background, i.e. in a very compact notation:  
\begin{equation}
\label{perturb_EOM}
E_i(\Phi_j^{(1)})=e^{H_i(\Box)}E_i^{(\ell)} 
(\Phi_j^{(1)}) = 0 
\, ,
\end{equation}
where $E(\Phi_j^{(1)})$ are the EoMs for the scalar and tensor perturbations $\Phi_j^{(1)}=\{\phi,\,h^{\perp}_{\mu\nu}\}$,
and $E^{(\ell)}$ is the local Starobinsky EoM in short notation. Notice that it appears an exponential  form factor for each perturbation. Therefore, the solutions for the linear perturbations are exactly the same than in the Starobinsky local theory. 

In order to get the EoM for perturbation in the form (\ref{perturb_EOM}), the form factors are uniquely fixed to \cite{Koshelev:2016xqb, SravanKumar:2018dlo}, 
  \be
&&     \gamma_{\rm S}(\square)=
     -\dfrac{1}{6}\dfrac{1}{{\square}+{R}/3}\cdot\left[e^{H_{0}({\square}+{R}/3)}\left(1-\dfrac{\square}{M^{2}}\right)-\left(1+\dfrac{{R}}{3M^{2}}\right)\right], \label{form_factor_S}  \\
&&     \gamma_{\rm C}(\square)=\left(\dfrac{{R}}{6M^{2}}+\dfrac{1}{2}\right)\left[\dfrac{e^{H_{2}(\square-2 R/3)}-1}{\square-2{R}/3}\right]. \label{form_factor_C}
\label{gammaC}  
\ee
We notice the Ricci scalar $R$ in $\gamma_{\rm C}(\square)$ (first term from the left) that introduces two more derivatives in the vertices in comparison to the kinetic term in Minkowski spacetime. As we will see such extra term destroys the renormalizability of the theory. 
The nonlocal theory is given by the action (\ref{NLtheory}) with the form factors 
(\ref{form_factor_S}), (\ref{form_factor_C}), while the entire 
functions $H_{0}(z)$ and $H_{2}(z)$ (see Appendix \ref{formfa}) are:

\begin{align} 
   H_{0}\left(\square+\dfrac{R}{3}\right)&=\dfrac{\alpha}{2}\{\ln p^{2}(\square)+\Gamma[0,\,p^{2}(\square)]+\gamma_{E}\} \, ,
   & 
   p(\square) &= \left(\square+\dfrac{R}{3}\right)^{n-3}(\square-M^{2}) \, , \label{H0} \\
  H_{2}\left(\square-\dfrac{2R}{3}\right)&=\dfrac{\alpha}{2}\{\ln q^{2}(\square)+\Gamma[0,\,q^{2}(\square)]+\gamma_{E}\} \, ,
   &   
   q(\square) &= \left(\square-\dfrac{2R}{3}\right)^{n-1}.
     \label{H2}
\end{align}
The theory \eqref{NLtheory} with form factors (\ref{form_factor_S}), (\ref{form_factor_C})
reduces to \eqref{localR^2} in the local limit $\Lambda_{*}\rightarrow\infty$. 

The degrees of the polynomials $p(z),\,q(z)$ differ by one in order to produce the same UV scaling in both terms of the (gauge-invariant part of the) propagator. Indeed \cite{Koshelev:2016xqb},  
\begin{equation}
\label{propagator}
    \mathcal{O}^{-1}(k)=-\frac{1}{k^{2}}\Big[\frac{P^{(2)}}{e^{H_{2}(-k^{2})}}-\frac{P^{(0)}}{(D-2)(1+\frac{k^{2}}{M^{2}})e^{H_{0}(-k^{2})}}\Big],
\end{equation}
where the first term is related to the tensor (spin-2) part, 
and the second one to the scalar part. As in local Starobinsky theory, the pole at $k^{2}=0$ corresponds to the graviton, while the pole at $k^2= - M^2$ corresponds to the scalaron or curvaton.

Summarizing, in this section we introduced a nonlocal model that has the same exact inflationary solution and the very same solutions for the linear perturbations as the Starobinsky theory \eqref{localR^2}. However, the presence of the Ricci scalar in (\ref{form_factor_C}) will be at the core of the renormalizability issue as we will show in the next section.

 \subsection{Power-counting analysis and the renormalizability problem} \label{Toy_model}
In order to study the renormalizability properties of the theory \eqref{NLtheory} with (\ref{form_factor_S}), (\ref{form_factor_C}), we consider a scalar higher-derivative toy model with the same UV behavior of the propagator and the same scaling in the leading vertices. 
For the power counting analysis it is sufficient to consider the asymptotic polynomial limit of the propagator \eqref{propagator}, namely 
\begin{equation}
\begin{split}
    \mathcal{O}^{-1}(k)&\sim-\frac{\Lambda_{*}^{2n-2}}{k^{2}}\left[\frac{P^{(2)}}{k^{2n-2}}-\frac{P^{(0)}}{\Lambda_{*}^2(D-2)(1+\frac{k^{2}}{M^{2}})k^{2n-4}}\right]\sim \frac{1}{\Lambda_{*}^2(k^2/\Lambda_{*}^2)^{n}} \, ,
    \end{split}
\end{equation}
which can be simulated through the following kinetic term for a scalar field $\phi$ (we here do not care about unitarity but only of the UV scaling of the loop integrals), 
\begin{equation}
  \mathcal{L}_{\rm K}\sim -\frac{1}{2}\phi\frac{\square^{n}}{\Lambda^{2n}}\phi \, \Lambda^{2},
\end{equation}
where $\Lambda \equiv \Lambda_{*}$ is the analog of the non-local mass scale.
Let us now focus on the vertices with most derivatives present in the action (\ref{NLtheory}) with \eqref{form_factor_S}, \eqref{form_factor_C},
\be
\label{vertices}
   && \mbox{from} \,\, \gamma_{\rm S} \, : \quad 
   \mathcal{R}\left[\dfrac{e^{H_{2}(\square-2R/3)}-1}{\square-2R/3}\right]\mathcal{R} \qquad \stackrel{\rm UV}{\rightarrow}  \quad \partial^{2} \square^{n-2} \partial^{2}\sim \Box^{n},  \\ 
\label{vertices2}
  && \mbox{from} \,\, \gamma_{\rm C} \, : \quad 
  \mathcal{R}\mathcal{R}\left[\dfrac{e^{H_{2}(\square-2R/3)}-1}{\square-2R/3}\right]\mathcal{R} \quad \stackrel{\rm UV}{\rightarrow} \quad  \partial^{4} \square^{n-2} \partial^{2}\sim\Box^{n+1}.
    \ee
The UV scaling of the above vertices can be obtained
by the following interaction vertex in the scalar field theory, 
\begin{equation}
\lambda_{2m}\phi^{2l}\dfrac{\square^{m}}{\Lambda^{2m}}\phi\,\Lambda^{2} \, , \quad l \in \mathbb{N} .
\end{equation}
$\lambda_{2m}$ is the coupling constant and we recover \eqref{vertices} or \eqref{vertices2} depending if we set $m=n$ or $m=n+1$.

In order to understand the contribution of the two vertices \eqref{vertices} to the renormalizability of \eqref{NLtheory}, we can then study the superficial degree of divergence of the Lagrangian
 \begin{equation}\label{toy2vert}
    \mathcal{L}= \Lambda^{2} \Bigg[-\frac{1}{2}\phi\frac{\square^{n}}{\Lambda^{2n}}\phi-\underbrace{\lambda_{2n}\phi^{2l}\frac{\square^{n}}{\Lambda^{2n}}\phi}_{\mathbb{V}_{2n}}-\underbrace{\lambda_{2n+2}\phi^{2l}\frac{\square^{n+1}}{\Lambda^{2n+2}}\phi}_{\mathbb{V}_{2n+2}}\Bigg] \, .
\end{equation}
It is worth stressing that $\mathbb{V}_{2n}$ is a vertex with at most $2n$ derivatives, while $\mathbb{V}_{2n+2}$ is a vertex with at most $2n+2$ derivatives. 

Therefore, the maximum superficial degree of divergence for a 1PI graph $\mathcal{G}$ in the scalar field theory \eqref{toy2vert} is obtained when all derivatives act on internal lines, namely 
\begin{equation}
    \omega(\mathcal{G}) \leqslant DL-2nI+2nV_{2n}+(2n+2)V_{2n+2},
    \label{omegaG}
\end{equation}
with $L$ loops, $I$ internal lines, and $V_{2n},\,V_{2n+2}$ count respectively the number of vertices with $2n$ or $2n+2$ derivatives. For the sake of simplicity, we define: 
\be
\gamma = n - \frac{D}{2} \, , 
\ee
and we replace the Euler's topological formula
\begin{equation}
    L=I-V_{2n}-V_{2n+2}+1 , 
\end{equation}
into (\ref{omegaG}). 
The final result reads:
\begin{equation}
      \omega(\mathcal{G})
    =D+2\gamma(1-L)+2V_{2n+2}.
\end{equation}
Let us look at the contribution of the two vertices separately. 

If $V_{2n+2}=0$, the superficial degree of divergence simplifies to:
\begin{equation}
\label{supdeg}
    \omega(\mathcal{G}) = D - 2 \gamma(L - 1).
\end{equation}
Hence, all diagrams are convergent for $L\ge 2$ whether we choose $\gamma  > D/2$, namely $n>D$. We still have divergences at $1$-loop in even dimension. In dimension $D=4$, 1-loop counterterms (in momentum space) show at most $\omega(\mathcal{G})=4$ powers in dimensional regularization and an arbitrary number $n_{E}\equiv E$ of external fields 
(see table \ref{table:1}).  
\begin{table}[t]
\footnotesize
\def\arraystretch{1.5}
\centering
\begin{tabular}{|c|c|c|c|}
\hline
\multicolumn{2}{|c|}{}
&\multicolumn{2}{|c|}{\text{1-loop Counterterms}} \\
\hline
\text{Vertices}&
\text{Max - divergence}&
\text{Scalar toy model} &\text{Non-local Gravity} \\
\hline
$\mathbb{V}_{2n}$ 
&$\omega(\mathcal{G}) =4$ & 
$\frac{1}{\displaystyle \varepsilon}\int d^4 x\,\phi^{n_E-1}\Box^2\phi$ 
& $\frac{1}{\displaystyle \varepsilon}\int d^4 x\sqrt{-g}\,\mathcal{R}^2$\\
\hline$\mathbb{V}_{2n+2}$ 
&
$\omega(\mathcal{G}) =4+V_{2n+2}$ &$\frac{1}{\displaystyle \varepsilon}\int d^4 x\,\phi^{n_E-1}\Box^{2+V_{2n+2}}\phi$&$\frac{1}{\displaystyle \varepsilon}\int d^{4}x\sqrt{-g}\,\mathcal{R}^{2+V_{2n+2}}$\\  
\hline
\end{tabular}
\caption{In the table, we display in a compact form the counterterms for a scalar higher-derivative theory and for the gravitational non-local theory. The integer $n_E$ stays for the number of external legs in the one-loop amplitude having any number of vertices $V_{2n+2}$ involved in the Feynman diagram. The presence of the 
$\Box^2$ (or $\Box^{2+V_{2n+2}}$) stays for one representative of all possible counterterms obtained by distributing the derivatives in all possible ways. 
In the gravitational theory such legs collect together in the correct curvature invariants. 
}
\label{table:1}
\end{table}
It deserves to be noticed that in the scalar toy model we have to add an infinite number of counterterms at one loop, namely the theory is actually not renormalizable. Conversely, in the gravitational theory and in absence of the vertices $V_{2n+2}$, we do not face this issue because the external graviton-legs collect into only few curvature invariants.

Let us move to the case where $V_{2n+2}\ne0$.
The superficial degree of divergence is
\begin{equation}
    \omega(\mathcal{G})=D - 2\gamma(L - 1)+2V_{2n+2}.
    \label{PCmale}
\end{equation}
Since the number of vertices $V_{2n+2}$ can be arbitrary, at one-loop $\omega(\mathcal{G})$ has no upper bound and an infinite number of counterterms must be introduced. 
Hence, the scalar toy model and nonlocal gravity will be both non-renormalizable. 
Indeed, the following counterterms (or analogous ones with less derivatives) have to be introduced consistently with (\ref{PCmale}) and with the number of external legs $E$ of the divergent diagram taken into account: 
\begin{equation}
    \frac{1}{\varepsilon}\int d^{4}x\sqrt{-g}\,\mathcal{R}^{E-j-1} \nabla^{4+2V_{2n+2}+2(j-E)}\mathcal{R},\quad V_{2n+2}\in\mathds{N}, \quad \max(0,E-2-V_{n+2})\le j\le E-2,
\end{equation}
where any possible permutation between covariant derivatives and curvatures is implied.
Since in the classical action \eqref{NLtheory} we have operators at most quadratic in the curvature, 
the theory is not renormalizable. 

We can summarize in short the result obtained for the scalar toy model \eqref{toy2vert} as follows: 
 \emph{if there are vertices with more derivatives than the kinetic operator $\mathcal{O}$, the renormalizability of the theory is destroyed.}
 
 On the contrary, a gravitational theory including vertices with as many derivatives as those in the kinetic operator (or less) will be \emph{super-renormalizable}, provided that the asymptotic polynomial in the form factor \eqref{form_factor_C} has large enough degree, 
 \begin{equation}
 \gamma > \frac{D}{2} \quad 
 \Longrightarrow \quad 
 \boxed{n>D} 
      \, .
 \end{equation}

 \subsection{Instabilities in a super-renormalizable non-local Starobinsky model}\label{stability}
According to the extensive discussion in the previous section, the operator \eqref{form_factor_C} spoils the renormalizability of the theory,
\begin{equation}
   \gamma_{\rm C}(\square)= \dfrac{{R}}{6M^{2}}\left[\dfrac{e^{H_{2}(\square-2 R/3)}-1}{\square-2{R}/3}\right]
   +
   \dfrac{1}{2} \left[\dfrac{e^{H_{2}(\square-2 R/3)}-1}{\square-2{R}/3}\right].
   \label{form_factor_C2}
\end{equation}
Indeed, the first nonlocal operator contributes to the vertices with two derivatives more than the other operators contributing to the propagator. 
Therefore, the simplest modification of the theory consists in removing such operator and study the implications for 
the stability. The theory is now super-renormalizable and the form factor 
$\gamma_{\rm C}$ reads:
 \begin{equation}\label{renormal_form_factor}
     \gamma_{\rm C}(\Box)=\frac{1}{2}\frac{1}{\Box-2R/3}[e^{H_2(\Box-2R/3)}-1] \, .
 \end{equation}
 However, as a drawback, the EoMs for the gravitational perturbations 
 turn into (see Appendix \ref{stab_problems}):
 \begin{align}
 \label{dSEOM}
     &\frac{1}{2}\left(\bar\square-\frac{\bar R}{6}\right)\left[e^{H_2\big(\bar\square-\frac{\bar R}{3}\big)}+\frac{\bar R}{3M^2}\right]h_{\mu\nu}^{\perp}=0 
     \stackrel{\text{(around dS)}}{\implies}\nonumber \\
&\frac{1}{2}\left(\bar\square-\frac{\bar R}{6}\right)\left[e^{H_{2}\left(\bar\square-\frac{\bar R}{3}\right)}+4\frac{H_{0}^{2}}{M^{2}}\right]h_{\mu\nu}^{\perp}=0 \, ,
 \end{align}
that are not in the form \eqref{perturb_EOM}, but an infinite number of complex massive zeros appears. 

In the paper \cite{Koshelev:2020fok} it was found that the stability condition for perturbations around a dS background metric is defined by the following parabolic region in the complex plane, 
\be
\label{Kosce0}
   [\Im(m_i^2)]^2 < 9H^2\Re(m_i^2) \, ,
\ee
where $m_i$ are related to the zeros $z_i$ of the following equation, 
  \be
    e^{H_2(z_i)}+C^2=0,\quad z_i=\frac{m_i^2-\bar R/3}{\Lambda_{*}^2}, \quad C^2=4\frac{H_{0}^{2}}{M^{2}} . 
\label{InfInst}
\ee
For numerical concreteness we recall that in our case the Hubble parameter at de Sitter inflation is estimated \cite{Koshelev:2016xqb} to be $H_{0}^{2}=\bar R/12$ with $\bar R\simeq240 M^2$, 
while $M=1.3\cdot 10^{-5}M_{\rm p}$, so that $C^2\simeq 240/3$.

Great Picard's Theorem ensures the algebraic equation (\ref{InfInst}) to be surely satisfied for some $z_i$ in open sets around the essential singularities of the analyzed function: in our case $\exp H(z)$ has essential singularities at infinity in several conical regions in the complex plane $z$.

In order to achieve the stability of the de Sitter solution, all the essential singularities of the candidate form factor $\exp H(z)$ should be enclosed in the region defined by (\ref{Kosce0}).
Unfortunately, all known form factors, purely exponential or asymptotically polynomial and studied till now, do not have such property. 
Therefore, we are not able to avoid an infinite number of instabilities. More details on the proof and plots about the distribution of zeros for the most relevant form factors are provided in Appendix \ref{stab_problems}.

Since to date, we do not have any explicit form factor fulfilling both the stability condition and the power counting renormalizability, in Section \ref{NLGM_approach}, we will evade the problem by proposing a new ultraviolet completion of the Starobinsky inflationary model based on a theory of gravity non-minimally coupled with matter.

Finally, let us give a physical interpretation of the result in comparison with a local theory. If the zeros had been finite in number, we could have estimated the background lifetime to be of the order of the inverse of the heaviest massive mode. Therefore, even if short, the lifetime would have been finite. In the case of the nonlocal theory, the number of runaway solutions is infinite. Therefore, the background lifetime is identically zero.
In other words, such background can not be realized in reality.

 \section{A super-renormalizable and stable non-local Starobinsky model}\label{NLGM_approach}
 In order to address the incompatibility of super-renormalizability and stability in the known proposals for a nonlocal extension of the Starobinsky theory, we here apply the recipe \cite{Modesto:2021ief} to the $f(R)$ gravity in the Einstein frame \eqref{Einstein_frame}. 
 The very same construction in the Jordan frame \eqref{fR}-\eqref{Jordan_frame} is problematic, as discussed in Appendix \ref{JF_details}.
 
The theory proposed in \cite{Modesto:2021ief} reads:

\begin{align}
&
\label{action}
\mathcal{S} [\Phi] = \int d^D x \sqrt{|g|} \left[ \mathcal{L}_{\text{loc}} + E_i F^{ij} (\hat{\De}) E_j \right]
,
\\
&
\mathcal{L}_{\text{loc}} = \frac{1}{\kappa^2} R  + \mathcal{L}_{\text{m}} (g_{\mu\nu}, \varphi, \psi, A^\mu)
,
\\
&
E_i (x) = \frac{\delta S_{\text{loc}}}{\delta \Phi^i (x)}
,
\label{EiEq1}
\\
&
\De_{ij} (x,y) = \frac{\delta E_{i} (x)}{\delta \Phi^j(y)} =  \frac{\delta^2 S_{\text{loc}}}{\delta \Phi^j(y) \delta \Phi^i (x)} = \hat{\De}_{ij} \, \delta^D (x,y)
,
\label{HessianEq1}
\\
&
2 \hat{\De}_{ik} \, F^{k} {}_j (\hat{\De}) = \left[ \frac{e^{{\rm H}(\hat{\De}_{\Lambda_{*}} )} }{e^{{\rm H}(0)} } - 1 \right]_{ij}, \label{FF2}
\end{align}
where $D$ is the number of spacetime topological dimensions, $\delta^D(x,y) = \delta^D (x-y)/\sqrt{|g(x)|}$ is the covariant delta function, $\Phi^i = (g_{\mu\nu}, \varphi, \psi, A^\mu)$
is the set of all fields (metric, scalars, fermions, gauge fields), $F^{ij}$ is a symmetric tensorial entire function whose argument is the Hessian operator $\hat{\De}_{ij}$, and ${\rm H}( \hat{\De}_{\Lambda_{*}} )$ is an entire analytic and asymptotically logarithmic function (see  Appendix \ref{formfa})\footnote{A wide class of such form factors is: 
\begin{eqnarray}
    H(z)=\alpha\int_0^{p(z)} d\omega\,\dfrac{1-f(\omega)}{\omega}, \quad f(\omega)=\exp({-\omega^n}).
\end{eqnarray} We already used one of this kind ($n=2$) in \eqref{general_form}.} whose argument is the dimensionless Hessian
\be
\label{DefStar}
( \hat{\De}_{\Lambda_{*}}) _{ij} = \frac{\hat{\De}_{ij}}{\left( \Lambda_{*} \right)^{[\hat{\De}_{ij}]}}\, .
\ee
In the above formula, $\Lambda_{*}$ is the non-locality mass scale and the dimensionality of the component $(i,j)$ of the Hessian is 
$[\hat{\De}_{ij}] = D - [\Phi^i] - [\Phi^j].$

 Moreover, if we apply the purely gravitational theory \eqref{fR} exhibits non-renormalizability as discussed in 
 
 As explain in the Appendix \ref{JF_details}, the direct implementation of the recipe in \cite{Modesto:2021ief} to 
the purely gravitational theory \eqref{fR} gives rise to a non-renormalizable theory. 
Therefore, our construction works only in the Einstein's frame.

 The $f(R)$ theory \eqref{Einstein_frame} is a tensor-scalar theory and, according to 
 \eqref{action}-\eqref{FF2}, its nonlocal completion reads:

\be
    S=\int d^4 x\sqrt{-g} \Bigg[
    \underbrace{\frac{M_{\rm p}^{2}}{2}R-\frac{1}{2}\partial_{\mu}\varphi\partial^{\mu}\varphi-V(\varphi)}_{\mathcal{L}_{\text{loc}}}+ \frac{1}{2}E_{i} \, ( e^{\rm{H}(\hat \Delta_{\Lambda_{*}})}-\mathds{1} )^{i}_{\, k}(\hat\Delta^{-1})^{kj}E_{j}\Bigg], 
    \label{NLGM_fR}
\ee

where $\Phi^{i}$ now stays 
for the graviton and the auxiliary scalar field (the scalaron or curvaton in ``$R^2$'' theory).

The special structure of the theory \eqref{NLGM_fR} is dictated by the following properties:

    (i) each solution of Einstein’s gravity is a solution of the NLGM theory with the same matter content;
    (ii) macro-causality is guaranteed;
    (iii) any exact solution of local Einstein’s theory coupled to matter, namely 
    \begin{equation}
        E^{(0)}(\Phi_{i}^{(0)})=0,
    \end{equation}
    is a solution of NLGM too.
    Moreover, the solution has the same stability properties in the nonlocal theory, i.e. 
  
    \begin{equation}
    \label{local_nonlocal_perturb}
        \boxed{\mathcal{E}_{k}^{(\ell)}(\Phi_{i}^{(\ell)}) = 0 \quad \Longrightarrow \quad  {{E}}_{k}^{(\ell)}(\Phi_{i}^{(\ell)})=0\,\,\quad \forall \ell\ge0 \quad \mbox{if} \quad E^{(0)}(\Phi_{i}^{(0)})=0 }
    \end{equation}
    with $E_{k}^{(\ell)}$ and ${\mathcal{E}}_{k}^{(\ell)}$ respectively representing the local and non-local EoMs at the $\ell^{th}$-order in the perturbations $\Phi_{i}^{(\ell)}$ of either the scalar or gravitational field. 
    In order to prove (\ref{local_nonlocal_perturb}), one can expand the fields as:
    \be 
    \Phi_i = \sum_{\ell = 0}^{+\infty} \epsilon^\ell \Phi_i^{(\ell)} \, .
    \ee
    Hence, the EoMs for the perturbations at the order $(\ell)$ reads:

      \begin{equation}
     \label{nonlocal_perturb}
{\mathcal{E}}_{k}^{(\ell)}=\sum_{m=0}^{\ell}\left[\left(e^{H^{(m)}(\hat\Delta_{\Lambda_{*}})}\right)_{kj}E_j^{(\ell-m)}+O\left(E^{(m)}E^{(\ell -m)}\right)\right]=0 \, . 
\end{equation}
    The above compact equation is solved according to (\ref{local_nonlocal_perturb}) at any perturbative order $(\ell)$; 
    (iv) the theory is super-renormalizable or finite at quantum level, and unitary at any perturbative order in the loop expansion. Renormalizability was first claimed in \cite{Modesto:2021ief} and then rigorously proved in \cite{Calcagni:2023goc}. In the following, we will extend the proof to a general potential $V(\varphi)$ as required by the inflationary scenario.

\subsection{Ultraviolet divergences}\label{renorm_proof}
In order to figure out the ultraviolet divergences and to study the renormalizability properties of the theory (\ref{NLGM_fR}), we can focus on the following higher derivative action,
\begin{equation}
\label{UV_action}
    S_{\rm UV}=\int d^4 x\sqrt{-g}\left[\frac{M_{\rm p}^{2}}{2}R-\frac{1}{2}\partial_{\mu}\varphi\partial^{\mu}\varphi-V(\varphi)+\alpha\sum_{k=0}^{n}E_{i}\frac{(\hat\Delta^{k}_{\Lambda_{*}})^{ij}}{\Lambda_{*}^{[\hat\Delta_{ij}]}}E_{j}\right] . 
\end{equation}
Indeed, only the asymptotic polynomial behaviour of the form factor can give rise to loop-divergences \cite{Calcagni:2023goc}, thus, the action (\ref{UV_action}) plays the role of a prototype for what concerns the divergences in a weakly, or quasi-polynomial, nonlocal theory. 
The dimensionless coefficient $\alpha$ yields $e^{\gamma_{\rm E}} e^{-{\rm H}(0)}/2$ as a consequence of the expansion of the form factor $F(z)$ for large $z$.

Let us move to a rough power-counting analysis of the loop amplitude for the theory (\ref{UV_action}). 
After having defined $N=n+2$, the propagator of scalar and gravitational fields for the theory (\ref{UV_action}) has the following scaling for large $k^2$, 
\begin{equation}\label{prop_scaling}
    \mathcal{O}^{-1}(k) \propto \frac{1}{k^{2N}} \, , \quad \mbox{in the ultraviolet regime} \, .
\end{equation}
Since the vertices have at most the same maximum number of derivatives shown in the kinetic operator, an upper bound to the superficial degree of divergence for an $L$-loops amplitude reads:
\be
\omega(\mathcal{G}) \le D L -  2 N I +  2 N V_{2 N} 
= D - (2N - D)(L-1) \, ,
\ee
where here $V_{2N}$ stays for the number of vertices with $2N$ derivatives, and we used the topological relation $L - 1 = I - V_{2N}$. 
Thus, if $N>D$ only one-loop diagrams can be ultraviolet divergent. In $D=4$ we require $N>4$ so that at most only one-loop divergences survive. Hence, the counterterms are all the possible operators with $4, 2, 0$ derivatives that can be constructed in a scalar-tensor theory. Such operators can be made explicit evaluating the square of the Starobinsky extremes, namely $E_i E_i$. 

In general, the exact power counting relation in $D=4$, assuming $N>4$ in order to have only one loop divergences, and taking into account all possible vertices, reads:
\be
\omega(\mathcal{G}) = 4 - 2 V_{2 N - 2} - 4 V_{2 N - 4} - d \, , 
\label{GPC}
\ee
where $V_{2 N - 2}$, $V_{2 N - 4}$ are the number of vertices with $2N-2$ and $2N-4$ derivatives respectively, while $d$ is the number of derivatives acting on external legs of the Feynman diagram. According to (\ref{GPC}) the diagram can contain any number of vertices with $2N$ derivatives.

\subsection{Renormalizability} \label{RENSE}
Given a general classical Lagrangian $\mathcal{L}$ whose action reads:
\be
S = \int d^D x \sqrt{|g|} \mathcal{L}\, , 
\label{classicG}
\ee
the renormalized Lagrangian (or action) $\mathcal{L}_{\rm ren}$ 
($S_{\rm ren} = \int d^D x \sqrt{|g|} \mathcal{L}_{\rm ren}$) is obtained starting from the classical Lagrangian 
$\mathcal{L}$ written in terms of the renormalized coupling constants $\alpha_i(t)$ ($t$ is related to the renormalization scale, see below) and adding the counterterms  (proportional to operators $\mathcal{O}_i$) to subtract the divergences. The counterterms are displayed by adding and subtracting the classical action in 
$\mathcal{L}_{\rm ren}$, namely 
\begin{align}
 \mathcal{L}_{\rm ren}     &=  
 \mathcal{L}(\alpha_i(t)) + \mathcal{L}_{\rm ct} 
  =   \mathcal{L}(\alpha_i(t)) + \sum_i \alpha_i(t) \left( Z_{\alpha_i} -1 \right) \mathcal{O}_i \nonumber\\
& =  \mathcal{L}\left(  Z_{\alpha_i(t)}  \alpha_i(t) \right) 
  =   \mathcal{L}(\alpha_{i, {\rm B} } )
 \, ,
 \label{LREN}
 \end{align}
where we introduced the following implicit definition of the renormalization scale $\mu$, 
\be
t := \frac{1}{(4 \pi)^2} \ln \frac{\mu}{\mu_0} \, . 
\ee 
 The last equality in (\ref{LREN}) makes explicit the renormalizability of the theory. Indeed, the Lagrangian functional is exactly the classical one but now depending on the bare coupling $\alpha_{i , {\rm B}}$. 
  Such an identification is valid if the operators $O_i$ are already present in the classical Lagrangian. However, if the latter property is not fulfilled, namely the theory is not renormalizable, we can still define the regularized Lagrangian as:
\be
\mathcal{L}_{\rm reg} = \mathcal{L} + \mathcal{L}_{\rm ct} \, .
\label{Lreg}
\ee
If the theory is not renormalizable, we have to compute an infinite number of loop corrections in order to define the regularized theory consistent with finite scattering amplitudes. Namely, such theory is undetermined because to compute and, in particular to measure, such an infinite number of terms we need an infinite amount of time. 
Notice that the regularized action can be computed only up to an infinite and undetermined number of front coefficients.  
On the contrary, in a renormalizable theory the renormalized and the classical action coincide in form. The latter property is shared by super-renormalizable nonlocal theories too. Nonlocal theories are ambiguous because we do not have a unique consistent theory, but once a classical action is chosen, namely a weakly nonlocal form factor is selected, only a finite number of operators gets renormalized.

In short, the regularized action can be always defined, but for the case of a renormalizable theory it coincides with the classical one, while for a non-renormalizable theory it consists in adding to the classical action an infinite number of new operators with undetermined coefficients. One could ask: is this really a problem? It is an issue if such procedure is ambiguous. 
Indeed, one could think to implement the following procedure. 
Given a non-renormalizable theory (for example the Einstein-Hilbert action), we can surely define the regularized 
Lagrangian according to (\ref{Lreg}) introducing an infinite number of counterterms; a finite set at each order in the perturbative expansion. Therefore, the physical quantum effective action, which we use at tree-level either to compute scattering amplitudes or to evaluate the effective EoM, will be perfectly finite. So far so good, but we can not forget that the finite contributions will depend on an infinite number of renormalization scales $\mu_i$. This is an equivalent way to see the infinite ambiguity inherent non-renormalizable theories. 
In the cut-off scheme, for a non-renormalizable theory we are forced to introduce an infinite number of Pauli-Villars fields, and, hence, an infinite number of scales $\mu_i$, one for each field \cite{Anselmi:1993cu}. 
In dimensional regularization the very same scale appear whether we identify the dimensionless cut-off $\epsilon$ with the dimensionful cut-off $\Lambda$, namely
\be
\ln \frac{ \Lambda^2 }{\mu_i^2} = \frac{1}{\epsilon_i} \, . 
\ee
The Lagrangian of counterterms is obtained computing the quantum effective action, which at one-loop  generically reads:
\be
&& \Gamma^{(1)} = S + \Gamma^{\rm div} + \Gamma^{\rm finite} \, , 
\ee
where the divergent contribution consists of a finite number of operators that we can express as:
\be
&& \Gamma^{\rm div} = - \frac{1}{2 (4 \pi)^2} \frac{1}{ \epsilon} \sum_i \beta_i \int d^D x \sqrt{|g|} \mathcal{O}_i \,. 
\label{epsilonLambda}
\ee
We here use the covariant cut-off regulator $L$  
\cite{bavi85}, which is related to the dimensional regularization one by 
\cite{bro-cass,bavi85}
\be
\ln L^2 \equiv \ln \frac{ \Lambda^2 }{\mu^2} = \frac{1}{\epsilon} = \frac{1}{2 - \omega}
=  \frac{2}{4-D} \, , \quad D = 2 \omega.
\label{dimreg}
\ee
where $\Lambda$ is the cut-off scale. Therefore, 
\be
\mathcal{L}_{\rm ct} = - \mathcal{L}_{\rm div} =  \frac{1}{2 (4 \pi)^2 } \frac{1}{ \epsilon} \sum_i \beta_i  \mathcal{O}_i
\equiv 
 \frac{1}{2 (4 \pi)^2} \ln \left( \frac{ \Lambda^2 }{\mu^2} \right) \, \sum_i \beta_i  \mathcal{O}_i
\, .
\label{LCT}
\ee
Comparing (\ref{LREN}) and (\ref{LCT}), we get:
\be
\alpha_i(t) \left( Z_{\alpha_i} -1  \right) 
= \frac{1}{2 (4 \pi)^2} \frac{1}{\epsilon}  \, \beta_i
= \frac{1}{2 (4 \pi)^2} \ln \left( \frac{ \Lambda^2 }{\mu^2} \right) \, \beta_i \, .
\label{aZ}
\ee
It is important to notice that in higher derivative local or weakly nonlocal theories with a large enough number of derivatives, the beta functions $\beta_i$ do not depend on the running coupling constants $\alpha_i(t)$. Hence, it is straightforward to solve (\ref{aZ}) consistently with $\alpha_i^{\rm B} \equiv \alpha_i Z_{\alpha_i}$, which is independent on the renormalization scale $\mu$, namely the product $\alpha_i Z_{\alpha_i}$ in independent on the renormalization scale although each factor changes with the scale, i.e. $\alpha_i(\mu)$ and $Z_{\alpha_i}(\mu)$. Hence, 
the product $\alpha_i Z_{\alpha_i}$ evaluated at different scales will take the same value, 
\be
[\alpha_i Z_{\alpha_i}] (\mu) =  [\alpha_i Z_{\alpha_i}(\mu_0)] \, .
\label{aZmuNo}
\ee
The condition of above, the equality (\ref{aZ}), and since $\beta_{\alpha_i}$ is independent on $\alpha_i(t)$, can be solved for $\alpha_i(\mu)$ as follows. 
Solving (\ref{aZ}) for $\alpha_i (\mu) Z_{\alpha_i} (\mu)$ and replacing the result in (\ref{aZmuNo}) we get:
\be
\alpha_i(\mu) + \frac{1}{2 (4 \pi)^2} \ln \left( \frac{ \Lambda^2 }{\mu^2} \right) \, \beta_i
= \alpha_i(\mu_0) + \frac{1}{2 (4 \pi)^2} \ln \left( \frac{ \Lambda^2 }{\mu_0^2} \right) \, \beta_i \, . 
\ee
The ultraviolet cut-off cancel out in the above equation, while the coupling constants are finite but change with the energy scale $\mu$ according to:
\be
\alpha_i(\mu) = \alpha_i(\mu_0) + \frac{\beta_{\alpha_i}}{(4 \pi)^2} \ln \frac{\mu}{\mu_0} \, .
\label{SimpSol}
\ee

Finally, the renormalized quantum effective action $ \Gamma^{{\rm eff}}_{\rm ren} $ is obtained computing the quantum effective action starting from (\ref{LREN}) instead of (\ref{classicG}), but at the same perturbative order in the loop expansion. The result is now finite, 
\be
 \Gamma^{{\rm eff}}_{\rm ren} =  S + \cancel{\Gamma^{\rm div}} + \Gamma^{\rm finite} + \cancel{\Gamma^{\rm ct}}
 =  S + \Gamma^{\rm finite} 
  \, . 
\ee

In general, including the case in which 
the beta functions depend on the running coupling constants, but the theory is renormalizable, it is worth noting that \cite{Shapirobook}:
\be
\frac{d \alpha_i}{d t} = \beta_{\alpha_i} \, . 
\label{RGEa}
\ee
The above renormalization group equations come from the renormalization group invariance of the renormalized quantum effective action, namely \cite{Shapirobook}:
\be
\frac{d \Gamma^{\rm eff}_{\rm ren}}{d \mu} =  0 \, \Longrightarrow \, 
\mu \frac{\partial}{\partial \mu} \Gamma^{\rm eff}_{\rm ren} + \frac{\beta_{\alpha_i}}{(4 \pi)^2} \frac{\partial}{\partial \alpha_i} \Gamma^{\rm eff}_{\rm ren} +  \sum_i \int d^D x \, \mu \frac{ d \Phi_i(x)}{d \mu} \frac{\delta}{\delta \Phi_i(x) }
\Gamma^{\rm eff}_{\rm ren}= 0 \, , 
\ee

which is solved by (\ref{RGEa}), and 
\be
\mu \frac{ d \Phi_i(x)}{d \mu} = \gamma_i \, \Phi_i(x) \, , 
\ee
which comes from the multiplicative renormalization of the fields (such term is not present for the metric).

All the computations are displayed in dimensional regularization, but the very same results can be obtained in any other regularization scheme including the cut-off scheme, if correctly implemented adding the Pauli-Villars fields \cite{Anselmi:1993cu}. 
An exemplification of the last statement is given in Appendix \ref{cutoff}.

Let us now come back to gravity. 
As a special case, in a purely gravitational theory in $D=4$, we have at most six divergent contributions in dimensional regularization, 
\be
\Gamma^{\rm div} = \frac{1}{2 \epsilon} \frac{1}{(4\pi)^{2}}\int d^4x \sqrt{-g} \left[  
\beta_{\rm GB} E + 
\beta_W C^2 + \beta_R R^2 + \beta_{\Box} \Box R+ \beta_G R + \beta_\Lambda 
\right] \, .
\ee
All the beta functions in a super-renormalizable theory with only one-loop divergences depend on the details of the form factor in the ultraviolet limit, but not on the six running coupling constants labelled by: 
$\alpha_{\rm GB}, \alpha_W, \alpha_R, \alpha_{\rm Ric}, \alpha_G, \alpha_{\Lambda}$. Therefore, for the whole set of couplings applies the result (\ref{SimpSol}). Indeed, the operators proportional to the latter couplings do not contain enough derivatives to contribute to the one-loop divergences. 

Conventionally, all the above operators have to be present in the classical theory in order to claim renormalizability. However, in super-renormalizable gravity with or without coupling to matter, the unitarity request may exclude some operators or fix the relative coefficients between them. As an example, for the simpler purely gravitational theory, i.e.
\begin{align}
\mathcal{L}_4 = - \lambda + \frac{1}{2 \kappa^2} \left[ R + G_{\mu\nu} \gamma(\Box) R^{\mu\nu} \right] = - \lambda + 
\frac{1}{2 \kappa^2} \left[R + R_{\mu\nu} \gamma(\Box) R^{\mu\nu} - \frac{1}{2} R\gamma(\Box) R  \right]  ,
\end{align}
the relative coefficient between the ${\rm Ric}^2$ and the $R^2$ operators is $-1/2$. Notice that the form factor is the same. Let us expand on this issue. We extract from the form factor the polynomial contribution, which is the only one to give rise to the ultraviolet divergences, and we add the $R^2$ and ${\rm Ric}^2$ operators needed for having renormalizability,
\begin{align}
\mathcal{L}_4 = - \lambda + \frac{1}{2 \kappa^2} R + \frac{1}{2 \tilde{\kappa}^2 } \left[ G_{\mu\nu} \left( \frac{e^{H(\Box)} - 1 - e^{\gamma_E} p(\Box) + e^{\gamma_E} p(\Box) }{\Box}  \right) R^{\mu\nu} \right] 
 + a R^2 + b R_{\mu\nu}  R^{\mu\nu}. 
 \label{splitAction}
\end{align}
All form factors consistent with power counting renormalizability are asymptotically polynomial, but here for the sake of simplicity we considered the Kuzmin's form factor \cite{Kuzmin}.

At most we have four running coupling constants, i.e. 
\be
\alpha_i(t) = \left\{ \kappa_4^{-2}(t), \lambda(t), a(t), b(t) \right\}.
\ee
At classical level, unitarity requires: $\kappa_4 = \tilde{\kappa}$, $a=b=0$, while at quantum level the very same values are imposed as initial conditions for the renormalization group equations (\ref{RGEa}), namely 
\be
\kappa_4(0) = \tilde{\kappa}\, , \quad a(0) = b(0) = 0 .
\ee 
Therefore, a non-zero value of the above running couplings due to the dependence on the renormalization scale $\mu$ (i.e. $\alpha_i(\mu) \sim \beta_{\alpha_i}  \log \mu/\mu_0$) can be reabsorbed in the logarithmic finite contributions also depending on $\mu$ (see eqs (\ref{GammaRen0}), (\ref{GammaRenEff}), which come later). Since the finite quantum corrections are perturbative, the extra Landau poles appearing in the resummed quantum propagator can not spoil unitarity because they are outside the perturbative approximation. In other words, the Landau poles appear exactly when the perturbative corrections are of the same order of magnitude of the classical contributions. 
Equivalently, the renormalization group invariance guarantees unitarity at any energy scale because 
the physical quantities are independent whether we use the pair $\alpha_i(\mu) , \mu$ or the pair $\alpha_i(\mu_0) , \mu_0$. 
Indeed, the quantum effective action can be expressed in terms of a renormalization invariant scale, say $\Lambda_{\rm RGI}$ such that $\partial_t \Lambda_{\rm RGI} = 0$. Again, if 
the quantum effective action is 
 the same for any value of the scale $\mu$, namely 
 \be
 \Gamma^{\rm eff}_{\rm ren}(\mu) = \Gamma^{\rm eff}_{\rm ren}(\mu_0), 
 \ee
 then, it can be evaluated at a specific point where its value is consistent with the absence of ghosts or other degrees of freedom.
Such scale is the experimental scale and the spectrum is the outcome of the observation. If only the graviton is observed at the scale $\mu$ then the RG-invariance guarantees that the spectrum consists only of the graviton at any energy scale and at any perturbative order. In a gravitational theory, the finite quantum correction to the kinetic operator is proportional to $k^2$, therefore, the residue at the pole of the graviton is unchanged by quantum corrections.  
It deserves to be noticed that if the classical theory was defined at a scale $\mu_0$ such that $a(\mu_0) \neq 0$ and $b(\mu_0) \neq 0$, then, the propagator would have acquired an infinite number of poles. 
Therefore, in order to secure unitarity, we assumed the
outcome of a measurement  at the scale $\mu_0$ to be only the graviton. 
Such choice is non-perturbative and defines the classical spectrum consistently with the experiment. At quantum level, the perturbative approach preserves the spectrum and unitarity is guaranteed according to the Cutkosky rules \cite{Briscese:2018oyx}.

We would like to rephrase the above renormalization group-invariant statement in a different way. Let us assume the theory to have divergences only at one loop that we can cancel with a finite number of counterterms proportional to operators not present in the classical action. Hence, a natural question reads: is the theory actually renormalizable? We are going to show that the quantum effective action is insensitive whether we start from a classical action with or without the two operators proportional to $a$ and $b$ in \eqref{splitAction}. 
For the sake of simplicity we do not introduce any cosmological constant. Moreover, we assume the form factor or, actually, the asymptotic polynomial only to give contributions to divergences proportional to the operators $R^2$ and Ric$^2$.

Therefore, \underline{ if $a$ and $b$ are present in the classical action} from the beginning, the quantum effective action before renormalization reads:
\begin{align}
 \Gamma^{(1)} = & \,\,\,S_4 + \Gamma^{\rm div} + \Gamma^{\rm finite} \nonumber \\
 = & \int  d^4 x \sqrt{-g} \Big\{ \frac{1}{2 \kappa^2} \left[ R + G_{\mu\nu} \gamma(\Box) R^{\mu\nu} \right]  
+  a(\mu) R^2 + b(\mu)  R_{\mu\nu}  R^{\mu\nu} \nonumber\\
& \hspace{2.5cm}- \frac{\beta_a}{2 (4 \pi)^2} \frac{1}{\epsilon} R^2 
- \frac{\beta_b}{2 (4 \pi)^2} \frac{1}{\epsilon} R_{\mu\nu} R^{\mu\nu} \nonumber \\
&\hspace{2.5cm} 
+ R\,  \Pi_a \left( - \frac{\Box}{\mu^2} ; \Lambda_* \right) R 
+  R_{\mu\nu} \,  \Pi_b \left( - \frac{\Box}{\mu^2}  ; \Lambda_* \right) R^{\mu\nu} \Big\} \nonumber 
\end{align}
\be && \hspace{0.55cm} =  
\int d^4 x \sqrt{-g} \Big\{ \frac{1}{2 \kappa^2} \left[ R + G_{\mu\nu} \gamma(\Box) R^{\mu\nu} \right]  
+  a(\mu) R^2 + b(\mu)  R_{\mu\nu}  R^{\mu\nu} \nonumber\\
&& \hspace{3.5cm} - \frac{\beta_a}{2 (4 \pi)^2} \frac{1}{\epsilon} R^2 
- \frac{\beta_b}{2 (4 \pi)^2} \frac{1}{\epsilon} R_{\mu\nu} R^{\mu\nu} \nonumber \\
&& \hspace{3.5cm} 
+ \frac{\beta_a}{2 (4 \pi)^2}  R\,  \log \left( - \frac{\Box}{\mu^2}  \right) R 
+  \frac{\beta_b}{2 (4 \pi)^2} R_{\mu\nu} \,  \log \left( - \frac{\Box}{\mu^2}  \right) R^{\mu\nu}  \nonumber \\
&& \hspace{3.5cm} 
+ R\,  \Pi^{*}_a \left( - \frac{\Box}{\Lambda_*^2}  \right) R 
+  R_{\mu\nu} \,  \Pi^{*}_b \left( - \frac{\Box}{\Lambda_*^2}  \right) R^{\mu\nu}
\Big\} \, , 
\ee
where we have decomposed the quantum form factor in two parts, the logarithmic part coming together with the divergence, thus depending on the renormalization scale $\mu$, and $\Pi^*_{a,b}$, which do not depend on the scale $\mu$. The latter form factor can show other non analytic contributions depending on the scale $\Lambda_*$. Indeed, the finite contributions very likely include $\log (- \Box)$ discontinuities because simply due to the Landau singularities. However, such operators will only depend on the non locality scale $\Lambda_*$.

The $1$-loop renormalized action is defined by:
\be
 \Gamma^{(1)}_{\rm ren}  =  S_4 + \Gamma^{\rm ct} \, , 
\ee 
 and the quantum effective action turns into:
 \be
 \Gamma^{(1){\rm eff}}_{\rm ren}& = & S_4 + \cancel{\Gamma^{\rm div}} + \Gamma^{\rm finite} + \cancel{\Gamma^{\rm ct}} \nonumber \\
 & = & 
\int d^4 x \sqrt{-g} \Big\{ \frac{1}{2 \kappa^2} \left[ R + G_{\mu\nu} \gamma(\Box) R^{\mu\nu} \right]  
+  a(\mu) R^2 + b(\mu)  R_{\mu\nu}  R^{\mu\nu}
 \nonumber \\
&& \hspace{2cm} 
+ \frac{\beta_a}{2 (4 \pi)^2}  R\,  \log \left( - \frac{\Box}{\mu^2}  \right) R 
+  \frac{\beta_b}{2 (4 \pi)^2} R_{\mu\nu} \,  \log \left( - \frac{\Box}{\mu^2}  \right) R^{\mu\nu}  \nonumber \\
&& \hspace{2cm} 
+ R\,  \Pi^{*}_a \left( - \frac{\Box}{\Lambda_*^2}  \right) R 
+  R_{\mu\nu} \,  \Pi^{*}_b \left( - \frac{\Box}{\Lambda_*^2}  \right) R^{\mu\nu}
\Big\} 
\nonumber \\
 & = & 
\int d^4 x \sqrt{-g} \Big\{ \frac{1}{2 \kappa^2} \left[ R + G_{\mu\nu} \gamma(\Box) R^{\mu\nu} \right]  \nonumber \\
&& \hspace{2cm} 
+  a(\mu_0) R^2 + \frac{\beta_a}{2 (4 \pi)^2} \log \left( \frac{\mu^2}{\mu_0^2} \right) R^2 \nonumber \\
&& \hspace{2cm} 
+ b(\mu_0)  R_{\mu\nu}  R^{\mu\nu} + \frac{\beta_b}{2 (4 \pi)^2} \log \left( \frac{\mu^2}{\mu_0^2} \right) R_{\mu\nu} R^{\mu\nu}  
 \nonumber \\
&& \hspace{2cm} 
+ \frac{\beta_a}{2 (4 \pi)^2}  R\,  \log \left( - \frac{\Box}{\mu^2}  \right) R 
+  \frac{\beta_b}{2 (4 \pi)^2} R_{\mu\nu} \,  \log \left( - \frac{\Box}{\mu^2}  \right) R^{\mu\nu}  \nonumber \\
&& \hspace{2cm} 
+ R\,  \Pi^{*}_a \left( - \frac{\Box}{\Lambda_*^2}  \right) R 
+  R_{\mu\nu} \,  \Pi^{*}_b \left( - \frac{\Box}{\Lambda_*^2}  \right) R^{\mu\nu}
\Big\} 
\, .
\label{GammaRen0}
\ee
{\em It is here crucial that the beta functions do not depend on $a$ and $b$, but only depend on the details of the polynomial in the form factor.  Therefore, the renormalization group invariance implies that $\Pi^*_{a,b}$ can not depend on $a$ and $b$ because the second and the third line of the last equality in (\ref{GammaRen0}) are already RG-invariant by themselves. Moreover, any further non analytic logarithmic contribution can only come accompanied by the non locality scale $\Lambda_*$}. 

Assuming $a(\mu_0) = b(\mu_0) = 0$ consistently with unitarity and/or the absence of ghosts in the classical theory, 
\be
&& \Gamma^{(1){\rm eff}}_{\rm ren} =  
\int d^4 x \sqrt{-g} \Big\{ \frac{1}{2 \kappa^2} \left[ R + G_{\mu\nu} \gamma(\Box) R^{\mu\nu} \right]  \nonumber \\
&& \hspace{3.3cm} 
+ \frac{\beta_a}{2 (4 \pi)^2}  R\,  \log \left( - \frac{\Box}{\mu_0^2}  \right) R 
+  \frac{\beta_b}{2 (4 \pi)^2} R_{\mu\nu} \,  \log \left( - \frac{\Box}{\mu_0^2}  \right) R^{\mu\nu}  \nonumber \\
&& \hspace{3.3cm} 
+ R\,  \Pi^{*}_a \left( - \frac{\Box}{\Lambda_*^2}  \right) R 
+  R_{\mu\nu} \,  \Pi^{*}_b \left( - \frac{\Box}{\Lambda_*^2}  \right) R^{\mu\nu}
\Big\} 
\, , 
\label{GammaRenEff}
\ee
where we identified $\mu_0$ with the renormalization group invariant scale. 
We can also assume $a(\mu_0), b(\mu_0) \neq 0$, but such initial conditions must be perturbative corrections to the classical action, namely two other RG-invariant scales have to be introduced and at which the spectrum includes only the graviton and no ghosts.

Let us start now \underline{without $a$ and $b$ in the classical action}, the quantum effective action before regularization reads:
\be
 \Gamma^{(1)}& = & S_4 + \Gamma^{\rm div} + \Gamma^{\rm finite} \nonumber \\
& = & \int d^4 x \sqrt{-g} \Big\{ \frac{1}{2 \kappa^2} \left[ R + G_{\mu\nu} \gamma(\Box) R^{\mu\nu} \right]  
- \frac{\beta_a}{2 (4 \pi)^2} \frac{1}{\epsilon} R^2 
- \frac{\beta_b}{2 (4 \pi)^2} \frac{1}{\epsilon} R_{\mu\nu} R^{\mu\nu} \nonumber \\
&& \hspace{2cm} 
+ R\,  \Pi_a \left( - \frac{\Box}{\mu^2} ; \Lambda_* \right) R 
+  R_{\mu\nu} \,  \Pi_b \left( - \frac{\Box}{\mu^2}  ; \Lambda_* \right) R^{\mu\nu}  \nonumber \\
& = & 
\int d^4 x \sqrt{-g} \Big\{ \frac{1}{2 \kappa^2} \left[ R + G_{\mu\nu} \gamma(\Box) R^{\mu\nu} \right]  
- \frac{\beta_a}{2 (4 \pi)^2} \frac{1}{\epsilon} R^2 
- \frac{\beta_b}{2 (4 \pi)^2} \frac{1}{\epsilon} R_{\mu\nu} R^{\mu\nu} \nonumber \\
&& \hspace{2cm} 
+ \frac{\beta_a}{2 (4 \pi)^2}  R\,  \log \left( - \frac{\Box}{\mu^2}  \right) R 
+  \frac{\beta_b}{2 (4 \pi)^2} R_{\mu\nu} \,  \log \left( - \frac{\Box}{\mu^2}  \right) R^{\mu\nu}  \nonumber \\
&& \hspace{2cm} 
+ R\,  \Pi^{*}_a \left( - \frac{\Box}{\Lambda_*^2}  \right) R 
+  R_{\mu\nu} \,  \Pi^{*}_b \left( - \frac{\Box}{\Lambda_*^2}  \right) R^{\mu\nu}
\Big\} \, , 
\label{withoutab}
\ee
where we decomposed the form factor in the logarithmic non analytic part depending on the renormalization scale $\mu$, plus the parts $\Pi^{*}_{a,b}$ depending only on the non locality scale $\Lambda_*$. The RG-scale $\mu$ here appears because of the logarithmic divergence. 

The regularized action is promptly obtained subtracting the divergent contributions to the classical one, namely
\be
 \Gamma^{(1)}_{\rm reg}& = & S_4 + \Gamma^{\rm ct} \, , 
\ee 
and the regularized quantum effective action reads:
\be
 \Gamma^{(1){\rm eff}}_{\rm reg}& = & S_4 + \cancel{\Gamma^{\rm div}} + \Gamma^{\rm finite} 
 + \cancel{\Gamma^{\rm ct}} \nonumber \\
 & = & 
\int d^4 x \sqrt{-g} \Big\{ \frac{1}{2 \kappa^2} \left[ R + G_{\mu\nu} \gamma(\Box) R^{\mu\nu} \right]  
 \nonumber \\
&& \hspace{2cm} 
+ \frac{\beta_a}{2 (4 \pi)^2}  R\,  \log \left( - \frac{\Box}{\mu^2}  \right) R 
+  \frac{\beta_b}{2 (4 \pi)^2} R_{\mu\nu} \,  \log \left( - \frac{\Box}{\mu^2}  \right) R^{\mu\nu}  \nonumber \\
&& \hspace{2cm} 
+ R\,  \Pi^{*}_a \left( - \frac{\Box}{\Lambda_*^2}  \right) R 
+  R_{\mu\nu} \,  \Pi^{*}_b \left( - \frac{\Box}{\Lambda_*^2}  \right) R^{\mu\nu}
\Big\} \, .
\label{GammaReg}
\ee
 Equivalently, we can replace $1/\epsilon$ according to (\ref{dimreg}) with $\log \Lambda^2/\mu^2$ into (\ref{withoutab}) and identify the cut-off $\Lambda$ with a renormalization group invariant scale, say $\mu$.

The quantum effective actions (\ref{GammaRenEff}) and (\ref{GammaReg}) are identical under the identification $\mu \equiv \mu_0$. This result is not very surprising because the theory is super-renormalizable, namely we have to add a finite number of counterterms. 
Notice that the very same proof does not apply to theories in which the beta functions depend on the running coupling constants.

\subsection{Finite number of counterterms}\label{finite_number}
The simple power counting analysis in (\ref{renorm_proof}), which is based on the asymptotic polynomial behaviour (see for example $H_0$ and $H_2$ in (\ref{general_form})), shows that with a proper choice of the degree of the polynomial $p(\Delta)$, we can avoid divergences in diagrams with a number of loops larger than one, namely we have divergences only at one loop. However, in order to avoid an infinite number of counterterms at one loop, the local foundational theory $S_\ell$ has to be consistent with the statements argued in \cite{Calcagni:2023goc}. In
short: {\em all the interaction terms, having the same number of derivatives of the kinetic terms, can only depend on derived matter fields, namely no matter field should appear without at least one derivative acting on it}.
As was already pointed out in  \cite{Calcagni:2023goc}, the above statement is enough to guarantee super-renormalizability even in presence of an analytic (but not necessarily polynomial) potential like for the Starobinsky model or a general $f(R)$ theory in the Einstein's frame. 

In \cite{Calcagni:2023goc} the latter statement was supported by the mean of scalar toy models.
In this section, we focus on the  theory  (\ref{UV_action}) and show that the number of operators contributing to the divergent part of the one-loop quantum effective action is finite regardless of the infinite number of monomials defining the potential $V(\varphi)=\sum_{n=0}^{\infty}a_n\varphi^n$ in $S_\ell$. Such operators will depend on $V(\varphi)$ and on a finite number of derivatives of the potential.

Let us investigate more quantitatively the proposed ultraviolet completion of the Starobinsky model. 

Since the divergent contributions to the quantum effective action only show up at one-loop (see Section (\ref{renorm_proof}), we can focus on such functional resulting from the path integral integration on the quantum fluctuation $h_{\mu\nu}$ around the background $\bar{g}_{\mu\nu}$, namely $g_{\mu\nu} = \bar{g}_{\mu\nu} + h_{\mu\nu}$, 
\begin{align}
      \Gamma^{(1)} &= S +\frac{i}{2}\mathrm{Tr}\ln{ H_{ij} } + \rm{g.f.} + {\rm g.h.} \, ,  
    \label{QEA} \\
 H_{ij}(x,y) &=  H_{ij}(x) \delta^D(x,y) = 
        \begin{pmatrix}
          H_{\alpha\beta,\rho\sigma}(x) & H_{\varphi,\rho\sigma}(x)  \\
          H_{\alpha\beta,\varphi}(x)  & H_{\varphi\varphi}(x)
        \end{pmatrix} \delta^D(x,y) 
      \nonumber \\  & \hspace{2.93cm}= 
       \begin{pmatrix}
          \dfrac{\delta^2 S}{\delta g_{\alpha\beta}\delta g_{\rho\sigma}} & \dfrac{\delta^2 S}{\delta\varphi\delta g_{\rho\sigma}}  \\ \\
          \dfrac{\delta^2 S}{\delta g_{\alpha\beta}\delta\varphi}  & \dfrac{\delta^2 S}{\delta\varphi\delta\varphi}
        \end{pmatrix} 
        \delta^D(x,y) \, ,
  \label{effect_action}
 \end{align}
    where $S$ is the non-local classical action evaluated on the background classical fields 
    ($\bar{g}_{\mu\nu}, \bar{\varphi}$), $H$ is the Hessian of the theory, and $\rm{g.f.}$ and ${\rm g.h.}$ stay for the gauge fixing and BRST ghost terms. 
    
    It is useful to rewrite the manifestly covariant Hessian expanding around the flat spacetime only the covariant derivatives acting on the fluctuation $h_{\mu\nu}$, namely for such operators we define $\bar{g}_{\mu\nu} = \eta_{\mu\nu} + f_{\mu \nu}$, where $f_{\mu\nu}$ is a classical perturbation that keeps track of the background metric (see for example \cite{JulveTonin}). Hence, the Hessian will consist of the free kinetic terms, manifestly covariant vertices, and vertices depending on $f_{\mu \nu}$ \cite{JulveTonin}. In a very short notation we can write:
 \be
 H=\mathcal{O}_0 + H_{\text{int}}(f_{\mu\nu} , R_{\rho\sigma} , \varphi, ...) . 
 \label{OHint}
 \ee

    Each element of $H$ is a function of the position, hence, the trace in (\ref{QEA}) consists in a sum on the indices $i,j$ and  an integral over the position space. In general \footnote{To prove (\ref{Trace}), we introduce in the first integral expression the completeness relation in the Fourier space (according to the expansion briefly explained above, the spacetime is locally Minkowski) twice,
    \be
    {\rm Tr} A(x,y) & = & \int d^D x \, \int d^D k_1 \int d^Dk_2 \, \langle x| k_1 \rangle A(x, \partial_x) \langle k_2 | x \rangle \langle k_1 | k_2 \rangle 
    \nonumber \\
&= &      \int d^D x  \int d^D k_1 \int d^Dk_2 \, e^{i x (k_1 - k_2)}   A(x, - i k_2) \delta^D (k_1 - k_2) \nonumber \\
& = &  \int d^D x \, \int d^D k_1 \,  A(x, - i k_1) \equiv  \int d^D x \, \int d^D k \,  A(x, - i k) \nonumber \\
& = &  \int d^D x \,  A(x, \partial_x ) \int d^D k \, e^{- i k (x - y) } \Big|_{x = y} 
    = \int d^D x \,  A(x, \partial_x)\, \delta^D(x-y)\Big|_{x = y} \, . \nonumber
  \label{TraceP}
    \ee}:
    \be
  {\rm Tr} A(x,y) = \int d^D x \, \langle x | A(x, \partial_x) | x \rangle = \int d^D x \,  A(x, \partial_x)\, \delta^D(x-y)\Big|_{x = y} \, .
  \label{Trace}
    \ee
 
  For simplicity, we omit $\sqrt{-g}$ in the denominator of each Dirac's delta and in the integrals' measure.
  Notice that the bra $\langle x |$ and ket $| x \rangle$ have dimension $M^{D/2}$, while the last equality is simply the generalization to a continuum index of the trace on a discrete space. Indeed, in a discrete space:
  \begin{align}
   \hspace{-1cm} {\rm Tr} A =   \sum_{i} \sum_j A_{i,j} \delta_{i,j} = \sum_i A_{ii} , 
 \end{align}
 which we have to compare with the analog trace in a continuum space:
 \begin{align}
   \hspace{-1.2cm}  \int \! d^D x \! \int \!  d^D y \, A(x,y)\, \delta^D(y-x) &= \int \! d^D x \int \! d^D y \, \left[  A(x, \partial_x) \delta^D(x - y) \right] \, \delta^D(y-x) \nonumber\\
 &= \int \! d^D x \,  A(x, \partial_x)\, \delta^D(x-y)\Big|_{x = y}\!\!\!\!\!  .
  \end{align}

 In order to extract the divergent part of the quantum effective action, we make the Taylor expansion of the trace in quantum effective action (\ref{QEA}), but only keeping the divergent contributions. Therefore, 
  \begin{align}
\Gamma^{\text{div}}& = \frac{i}{2}\mathrm{Tr}\ln H\Big|_{\rm div} 
= \frac{i}{2} \mathrm{Tr}\ln\left[\mathcal{O}_0\right]+\frac{i}{2}\mathrm{Tr}\ln\left[1+\mathcal{O}_0^{-1}H_{\text{int}}\right]\Big|_{\rm div}\nonumber\\ 
&= \frac{i}{2}\sum_{n=1}^{\infty}\dfrac{(-1)^{(n+1)}}{n}\mathrm{Tr}\left[(\mathcal{O}_0^{-1}H_{\text{int}})^n\right]\Big|_{\rm div} .
\label{DivTrun}
\end{align}

Moreover, since we look for the divergent part of the quantum effective action, it is sufficient to consider the prototype model (\ref{UV_action}), and in particular the limit of large Euclidean momentum squared. In such limit, the Hessian takes contributions only from the operators quadratic in $E_i$ and reads:
\be
H^{\rm UV}_{i j}  = \frac{ \delta^2 S_{\rm UV  }}{\delta\Phi_i\delta\Phi_j}= \frac{\alpha}{\Lambda_{*}^{[\hat\Delta_{ij}]}} \dfrac{\delta^2( \sqrt{|g|} E_i \sum_k(\hat\Delta^k_{\Lambda_{*}})^{ij} E_j)} {\delta\Phi_i\delta\Phi_j}\,  , \quad 
\Phi_i \in \{ g_{\mu\nu}, \varphi\} \, .
\ee
In order to simplify our analysis, we redefine the scalar field so as to become dimensionless as well as the graviton, namely $\varphi \rightarrow M_{\rm p} \, \varphi$, thus $E_\varphi \rightarrow M_{\rm p}^2 E_\varphi$ (see the Appendix \ref{ADetails}). For the sake of simplicity, we also assume $\Lambda_{*} \equiv M_{\rm p}$. 
Hence, in terms of the new EoMs \eqref{EOM_Hessian}--\eqref{EOM_grav}, the action (\ref{UV_action}) turns into:
\begin{align}
   S_{\rm UV}= M_{\rm p}^2 \int d^4 x\sqrt{-g} \left[
 \frac{1}{2}R-\frac{1}{2}\partial_{\mu}\varphi\partial^{\mu}\varphi- \frac{V( M_{\rm p}\varphi)}{M_{\rm p}^2} 
 + \frac{\alpha}{M_{\rm p}^2} \sum_{k=0}^{n} 
 \, 
 M_{\rm p}^2 E^\prime_{i} \, 
 \frac{(\hat{\Delta}^{\prime k} M_{\rm p}^{2 k} )^{ij}   }{M_{\rm p}^{4 k} M_{\rm p}^{4} } 
 \, 
 M_{\rm p}^2 E^\prime_{j}\right], 
    \label{NLGM_fRMM0}
\end{align}
 
where all the fields are now dimensionless (see Appendix \ref{ADetails}).  
We further simplify the above action introducing a new notation for the potential as a function of the dimensionless scalar field, 
\be
\mathcal{V}(\varphi) \equiv \frac{V( M_{\rm p}\varphi)}{M_{\rm p}^2} \, , \quad  [\mathcal{V}(\varphi)] = 2 \, .
\ee
Finally, the action reads:
\be
   S_{\rm UV}= M_{\rm p}^2 \int d^4 x\sqrt{-g} \left[
 \frac{1}{2}R-\frac{1}{2}\partial_{\mu}\varphi\partial^{\mu}\varphi- \mathcal{V}(\varphi)
 +\frac{\alpha}{M_{\rm p}^2} \sum_{k=0}^{n} E^\prime_{i} \left( \frac{\Delta^{\prime k}}{M_{\rm p}^{2k} } \right)^{ij} \!\!E^\prime_{j}\right], 
    \label{NLGM_fRMM}
\ee

  where $E^\prime_{i}$ and $\Delta^\prime$ are defined in (\ref{ADetails}). 
  According to the power counting analysis in \eqref{GPC}, $\omega(\mathcal{G}) = 0,2,4$, thus, loop integrals are divergent only when are present vertices 
 
  having $2N=2n+4$, $2 {N}-2=2n+2$ and $2 {N}-4=2n$ derivatives, namely only
   $V_{2n+4}$, $V_{2n+2}$, and $V_{2n}$ have to show up in the Feynman integrals. 

 More precisely, according to \eqref{GPC}, one-loop diagrams with $V_{2N}$ arbitrary, $V_{2N-2}=2, V_{2N-4}=0$, or $V_{2N-2}=0, V_{2N-4}=1$ are the only power-counting divergent ones. Moreover, looking at \eqref{DivTrun}, it is clear that every term of the series corresponds to a particular one-loop divergent diagram, 
 involving the vertices coming from the interacting term in \eqref{OHint}. Furthermore, according to \cite{Calcagni:2023goc}, counterterms coming from $\omega=4$ diagrams cannot depend on 
 $\mathcal{V}(\varphi)$.

Following \cite{Asorey:1996hz, Shapirobook}, the Hessian in a convenient gauge takes the following minimal form, namely there are no vertices with $2n+4$ derivatives, 
  
\begin{align}
 H^{ij}_{\rm UV}  \propto  
\delta^{i j} \Box^{n+2} 
&+ \underbrace{{\bf V}^{i,j , i_1  \dots i_{2 n +2}} 
\nabla_{i_1}  \cdots \nabla_{i_{2 n+2 }} }_{\text{$\sim V_{2N-2}$ vertices}}+
 {\bf W}^{i j , i_1  \dots i_{2 n + 1}}  
\nabla_{i_1} \cdots \nabla_{i_{2 n+1}} \nonumber\\ 
 &+ \underbrace{{\bf U}^{i,j, i_1 \dots i_{2 n }
\nabla_{i_1}  \cdots \nabla_{i_{2 n}}}}_{\text{$\sim V_{2N-4}$ vertices}} +O(\nabla^{2n-1}) \, , 
\label{Hess2}
\end{align}
where proportional means that we have omitted an overall constant divided by $M_{\rm p}^{2n}$. 
The tensors ${\bf V}$, ${\bf W}$, and ${\bf U}$ are easy to guess, up to integrations by part that can increase (by a finite number) further the derivatives of $\mathcal{V}$, on the dimensionality ground of the operators present in the action (\ref{NLGM_fRMM}). For the sake of simplicity, we here consider only the monomial theory, namely in (\ref{NLGM_fRMM}) it is present only $\Delta^n$, hence:

\begin{align}
\hspace{-0.5cm}
[ {\bf V}^{i,j , i_1  \dots i_{2 n +2}}  ] = 2 \,\, &\Longrightarrow \,\, 
 {\bf V}  = \sum_i c_i^v \mathcal{O}^v_i 
  \, ,
 \label{VU}\\
 \hspace{-0.5cm}
  [ {\bf W}^{i,j , i_1  \dots i_{2 n +1}}  ] = 3 \,\, &\Longrightarrow \,\, 
 {\bf W} =  \sum_i c_i^w \mathcal{O}^w_i\, ,\label{WDOPPIO}\\
 \hspace{-0.5cm}
 [ {\bf U}^{i,j , i_1  \dots i_{2 n }}  ] = 4 \,\, &\Longrightarrow \,\, 
 {\bf U} = 
 \sum_i c_i^u \mathcal{O}^u_i, \label{VWU}
\end{align}
with
\begin{align}
\mathcal{O}_i^v \supset  \{&
 \partial^2 \Phi, \mathcal{V} ,  \mathcal{V}^{'} ,  \mathcal{V}^{''} ,  \, \mathcal{V}^{'''} , \, 
 \mathcal{V}^{''''} \}
 \times  \varphi^{(\rm R)}\,,\label{setOv}\\
   \mathcal{O}^w_i \supset \{&  \partial(\partial^2 \Phi) , \, \partial(\mathcal{V}) ,  \, \partial(\mathcal{V}') , \, \partial(\mathcal{V}^{''}) 
 \, \partial(\mathcal{V}^{'''}) \, , 
\partial( \mathcal{V}^{''''}) 
 \} 
 \times  \varphi^{(\rm R)}
 \, ,  \\
 \mathcal{O}^u_i \supset \{&
 (\partial^2 \Phi_i ) (\partial^2 \Phi_j), 
 \,  \mathcal{V}^2, \,   (\mathcal{V}')^2 , 
 \, (\mathcal{V}^{''})^2 , 
  \, (\mathcal{V}^{'''})^2 ,
   \, (\mathcal{V}^{''''})^2 ,
   \nonumber \label{setOu}\\
 & \mathcal{V}  \mathcal{V}^{'} ,   \mathcal{V}  \mathcal{V}^{''} ,   \mathcal{V}  \mathcal{V}^{'''}  , 
 \, \mathcal{V}  \mathcal{V}^{''''} , 
  \mathcal{V}^{'}  \mathcal{V}^{''} ,  \mathcal{V}^{'}  \mathcal{V}^{'''} ,   \mathcal{V}^{'}  \mathcal{V}^{''''}  , 
   \mathcal{V}^{''}  \mathcal{V}^{'''} ,   \mathcal{V}^{''}  \mathcal{V}^{''''} , 
  \mathcal{V}^{'''}  \mathcal{V}^{''''}
  \} \times  \varphi^{(\rm R)},
\end{align}
where by $\times \varphi^{(\rm R)}$ we very shortly mean other powers of $\varphi$ not coming from the potential
that can only multiply the other operators because dimensionless. The other powers of $\varphi$, i.e. 
$\varphi^{(\rm R)}$, come from extra terms in $E_i$ and $\Delta$ beside the potential. Since the latter terms show up in a finite number, the total number of operators in the tensors ${\bf V}, {\bf W}, {\bf U}$ is strictly finite. This is the crucial evidence in support of a finite number of counterterms. 
The above analysis is general, but for our theory ${\bf W}$ is actually not present because in the classical action there are no boundary terms. 
The case with $\Delta^{n-1}$ and $\Delta^{n-2}$ present in action (\ref{NLGM_fRMM}) implies the presence of extra operators in the set (\ref{setOv}) with $M_{\rm p}^2$ in place of $\partial^2$, i.e.
\be
\mathcal{O}_i^v \supset  \{
 \partial^2 \Phi, 
M_{\rm p}^2 \Phi,
 \mathcal{V} ,  \mathcal{V}^{'} ,  \mathcal{V}^{''} ,  \, \mathcal{V}^{'''} , \, 
 \mathcal{V}^{''''} \}
 \times  \varphi^{(\rm R)}
  \, ,
\ee
and extra operators in the set (\ref{setOu}) with $M_{\rm p}^2$ in place of $\partial^2$ or $M_{\rm p}^4$ in place of $\partial^2 \partial^2$, namely 
\begin{align}
 \mathcal{O}^u_i \supset \{&
 (\partial^2 \Phi_i ) (\partial^2 \Phi_j), 
 (M^2_{\rm p} \Phi_i ) (\partial^2 \Phi_j),
 (M^4_{\rm p} \Phi_i ) ,
 \,  \mathcal{V}^2, \,   (\mathcal{V}')^2 , 
 \, (\mathcal{V}^{''})^2 , 
  \, (\mathcal{V}^{'''})^2 ,
   \, (\mathcal{V}^{''''})^2 ,
   \nonumber \\
   &
  \mathcal{V}  \mathcal{V}^{'} , \,  \mathcal{V}  \mathcal{V}^{''} , \,  \mathcal{V}  \mathcal{V}^{'''}  , 
 \, \mathcal{V}  \mathcal{V}^{''''} , 
\,   \mathcal{V}^{'}  \mathcal{V}^{''} , \,  \mathcal{V}^{'}  \mathcal{V}^{'''} , \,  \mathcal{V}^{'}  \mathcal{V}^{''''}  , 
\,   \mathcal{V}^{''}  \mathcal{V}^{'''} , \,  \mathcal{V}^{''}  \mathcal{V}^{''''} , 
\,  \mathcal{V}^{'''}  \mathcal{V}^{''''}
  \} \times  \varphi^{(\rm R)}
   \, . 
\end{align}

Therefore, we can display the divergent contributions to the trace collecting $\Box^{n+2}$ in front and making the Taylor expansion of the $\log-$function, 
\begin{align}
  {\rm Tr} \left[  \log H^{ij}_{\rm UV}  \right] & =  
 (n+2)  {\rm Tr}\left[  \log \delta^{ij} \Box \right] \nonumber \\
 & + {\rm Tr} \left[  {\bf V}^{i,j , i_1  \dots i_{2 n +2}} 
\nabla_{i_1}  \cdots \nabla_{i_{2 n+2 }} \, \frac{1}{ \Box^{n+2} } \right] \nonumber \\
&
+ {\rm Tr} \left[ {\bf W}^{i j , i_1  \dots i_{2 n + 1}}  \nabla_{i_1} \cdots \nabla_{i_{2 n+1}}  \,  \frac{1}{ \Box^{n+2} }
\right] 
\nonumber 
\\
&
+ {\rm Tr} \left[  
 {\bf U}^{i,j, i_1 \dots i_{2 n }} 
\nabla_{i_1}  \cdots \nabla_{i_{2 n}} \,   \frac{1}{ \Box^{n+2} } 
 \right] 
\nonumber 
 \\
&- \frac{1}{2} 
{\rm Tr} \left[  \delta_{lm}
 {\bf V}^{i,l , i_1  \dots i_{2 n +2}} 
\nabla_{i_1}  \cdots \nabla_{i_{2 n+2 }} \, \frac{1}{ \Box^{n+2} } 
 {\bf V}^{m,j , i_1  \dots i_{2 n +2}} 
\nabla_{i_1}  \cdots \nabla_{i_{2 n+2 }} \, \frac{1}{ \Box^{n+2} }
\right] . 
\label{acca}
\end{align}
We can now implement the trace (\ref{Trace}) in (\ref{acca}) with the aim of list all possible counterterms. Again we focus on the monomial theory.

The trace of the first operator, i.e $\log \Box^{n+2}$, can only give purely gravitational divergences with four derivatives. In short:
\be
{\rm Tr} \log \Box \quad \Longrightarrow \quad 
\sqrt{-g} {\mathcal R}^2 , \,\,  \sqrt{-g}  \Box { R}  \, . 
\ee
where by $\mathcal{R}$ we here mean the Riemann tensor, the Ricci tensor, or the Ricci scalar. 

   The trace of the second operator in (\ref{acca}) reads:
   \be
  \int d^4x \,  {\bf V} \, 
\nabla^{2n+2} \, \frac{1}{ \Box^{n+2} } \, \delta^4(x- y)\Big|_{x = y} 
\quad \Longrightarrow \quad``\,\nabla^2 {\mathcal{O}}^v_i\," \, ,
   \ee
   which gives rise to counterterms with two or four derivatives. Operators with four derivatives are easy to display, 
  
   \begin{align}
\label{O4nonpoly}
   & \frac{1}{\varepsilon}\sqrt{-g}R_{\mu\nu}R^{\mu\nu}, && \frac{1}{\varepsilon}\sqrt{-g}R^{2}, && \frac{1}{\varepsilon}\sqrt{-g}(\square\varphi)^{2}, &\frac{1}{\varepsilon}\sqrt{-g}(\nabla\varphi)^{2}\square\varphi, \nonumber\\
   &\frac{1}{\varepsilon}\sqrt{-g}(\nabla\varphi)^{4}, &&\frac{1}{\varepsilon}\sqrt{-g}R(\nabla\varphi)^{2}, &&\frac{1}{\varepsilon}\sqrt{-g}R^{\mu\nu}(\nabla_{\mu}\varphi)(\nabla_{\nu}\varphi) , &\frac{1}{\varepsilon}\sqrt{-g}R\square\varphi \, ,
\end{align}
while for the case of two derivatives we provide an implicit form,
\be
 \quad \frac{1}{\varepsilon}\sqrt{-g} \sum_i c_i^v \partial^2 \mathcal{O}^v_i(\varphi) \, , 
\label{palle}
\ee
where the operators ${\mathcal{O}}^u_i$ in (\ref{palle}) are extracted from the set (\ref{VU}), but do not carry any derivative, namely only a subset of the terms in (\ref{VU}) can appear in (\ref{palle}). In particular, if the two derivatives in (\ref{palle}) act on the metric tensor, the divergent contributions look like:
\be
\frac{1}{\varepsilon}\sqrt{-g} \sum_i c_i^v R \, {\mathcal{O}}^u_i \, .
\ee

  The trace of the third operator in (\ref{acca}) is not relevant for our theory. Therefore, 
   \be
  \int d^4x \,  {\bf W } \, \nabla^{2n+1} 
 \, \frac{1}{ \Box^{n+2} } \, \delta^4(x- y)\Big|_{x = y} 
\quad \Longrightarrow \quad ``\,\nabla {\mathcal{O}}^w_i = 0 \," \, .
   \ee
      From the trace of the fourth operator in (\ref{acca}) we get:
   \be
  \int d^4x \,  {\bf U } \nabla^{2n} 
 \, \frac{1}{ \Box^{n+2} } \, \delta^4(x- y)\Big|_{x = y} 
\quad \Longrightarrow \quad ``\, {\mathcal{O}}^u_i \," \, , 
   \ee
   Therefore, we get again divergences with four derivatives or zero derivatives, but not two derivatives.

 Finally, the trace of the fifth operator in (\ref{acca}) reads:
   \be
  \int d^4x \,  {\bf V \, V } 
  \,
  \nabla^{2n+2} 
 \frac{1}{ \Box^{n+2} } \, 
 \nabla^{2n+2} 
 \, \frac{1}{ \Box^{n+2} } \, 
 \delta^4(x- y)\Big|_{x = y} 
\quad \Longrightarrow \quad ``\, ( {\mathcal{O}}^v_i)^2\," \, .
   \ee
   which implies counterterms with two, four, and zero derivatives that we briefly write:
     \begin{equation}\label{O0}
        \frac{1}{\varepsilon}\sqrt{-g}
        \sum_j c_{ij}^v\mathcal{O}^v_i\mathcal{O}^v_j .
    \end{equation}
   The explicit form of the above divergences with four derivative falls into the class displayed in (\ref{O4nonpoly}).

   We conclude by emphasizing once again that the purpose of this section has been to show that the number of divergent one-loop integrals is finite despite the analytic potential hidden in a general $f(R)$ theory in the Einstein's frame. Therefore, the theory requires a finite number of one-loop counterterms, while there are no divergences for $L>1$.

\subsection{Achieving finiteness}\label{ways_to_SR}
The previous section secured the presence of a finite number of divergences, thus, the theory is 
super-renormalizable according to the rigorous analysis in the spectrum consists at most of the graviton and the scalar curvaton action (\ref{RENSE}).
However, we can go beyond and make the theory finite with a simple modification of the form factor. In particular, we can replace the polynomial of $\hat{\Delta}_{\Lambda_*}$ in the exponential form factor (\ref{NLGM_fR}) with 
\cite{Calcagni:2023goc}:

  \begin{equation}
        p(\hat{\Delta}_{\Lambda_*})_{ij} = \hat\Delta \left[ \tilde a_{n+1} \tilde{\Delta}^n + \tilde a_{n} \tilde{\Delta}^{n-1} + \dots + \tilde a_1 + \left(  \sum_r c_r O_r  \right) \mathds{1}\Box^{n-2} \right]_{ij}\,,
        \label{PolyKiller}
    \end{equation}
    where by $O_r$ we mean the very same operators in the counterterms \eqref{O4nonpoly}--\eqref{O0}.
\\   
    It has been extensively shown in 
    \cite{Calcagni:2023goc} that the other operators in (\ref{PolyKiller}) of the form:
    \be
    E_i \left( {\mathcal O}_r \Box^{n-2} \right)_{ij} E_j
    \label{mimetic}
    \ee 
    give rise to divergences of the same kind displayed in the previous section, but now linear in the front coefficients $c_r$. Hence, we can always make zero all the one-loop beta functions with a proper choice of the constant coefficients $c_r$ and achieve finiteness. This result is not a fine tuning but one-loop exact because there are no divergences for $L>1$.
    Last but not least, such so called mimetic killers $\mathcal{O}_r$ (\ref{mimetic}) have been introduced inside the form factor in order to not spoil the stability properties of the classical theory. 
\section{Conclusions}
The outcome of this paper is twofold. 
In the effective theory approach, we have shown, by comparing the $R^2$ operator with the higher derivative operators $\mathcal{O}(\mathcal{R}^3)$, that the Starobinsky theory can not be an effective theory. Hence, since $R+R^2$ is non-renormalizable worse than the Einstein-Hilbert one, the  ultraviolet completion of the Starobinsky action is unavoidable. 
Therefore, we made a review of the ultraviolet completion of the Starobinsky model in the nonlocal gravity framework \cite{Koshelev:2016xqb, SravanKumar:2018dlo}, and we pointed out some issues. In particular, we showed that the model proposed, developed, and extensively studied by authoritative scientists in the past, is not able to achieve at the same time the stability and the renormalizability. Indeed, if the theory is super-renormalizable the infinite number of derivatives gives rise to an infinite number of instabilities (run-away solutions) that make identically zero the lifetime of the Starobinsky inflationary solution. In other words, the Starobinsky inflation does not take place. On the other hand, if we insist in having stability we end up with a non-renormalizable theory more divergent than the original Starobinsky theory. 

In order to overcome the above inconsistency, we have proposed a new nonlocal ultraviolet completion of the Starobinsky theory that shares with the latter all the classical properties, namely: same exact solutions, same stability properties, macro-causality. Moreover, the new proposal, based on the general action introduced in \cite{Modesto:2021ief} and extensively studied in \cite{Calcagni:2023goc}, is at least super-renormalizable or even finite at quantum level depending on the kind of local limit \cite{Anselmi:2024ocm, Anselmi:1991wb, Anselmi:1992hv, Anselmi:1993cu} in the ultraviolet regime.

Last but not least, the reason that motivated us to write this paper, but also previous papers by us and other authors, is related to the formidable success of the inflationary paradigm, namely to the following question: why can we explain so well the physics of the Universe from very short to very large distances without deeply getting inside the quantum gravity regime?
According to the outcome of this paper, any $f(R)$ theory is a trustable theory till very high energy because it exists an ultraviolet completion that preserves the physical solutions and stability properties of the $f(R)$ model itself. Such ultraviolet completion is defined by a finite and unitary nonlocal theory of quantum gravity and matter in the very conservative quantum field theory framework. 
Perturbative quantum gravity can only provide small deviations from the predictions of the inflationary specific model. 
In other words, all the results of the proper inflationary model are exact solutions of the fundamental theory too.

\appendix
\section{Form Factors}\label{formfa}
In this section we review two classes of form factors extensively implemented in quantum field theories:
asymptotically polynomial form factors and asymptotically exponential form factors. 
\subsection{Asymptotically polynomial form factors}
A general class of entire functions compatible with unitarity, but in particular with the power counting super-renormalizability \cite{Kuzmin,modesto}
can be defined through the following integral, 
\be
\label{Hpoly}
{\rm H}(x)\equiv \alpha \int_0^{p(x)}  \frac{1-f(z)}{z} \, dz \,, 
\ee 
where $p(x)$ is a generic polynomial of $x$ and degree $n$, while $\alpha$ is a positive constant.
In order to achieve an asymptotic polynomial behavior of $\exp \mathrm H(x)$ as requested from the power counting renormalizability \cite{Kuzmin,modesto}, $p(x)$ and $f(z)$ in \eqref{Hpoly} must satisfy the following requirements~\cite{Kuzmin,modesto}: 
\begin{enumerate}
\item $p(x)$ is a real polynomial of degree $n$ such that  $p(0) = 0$, 
\item $f(z)$ is an entire and real function on the real axis with $f(0) = 1$,
\item  $|f(z)| \rightarrow 0$ for $|z| \rightarrow + \infty$ in a conical region around the real axis. 
\end{enumerate}

As an example, let us consider the following function 
\be
\label{eq.gz} f(z) = \exp(-z^n)\, .
\ee   
In this case, the result of the integral (\ref{Hpoly}) gives the following entire function,
\be
\label{Hx}
{\rm H}(x)=\frac{\alpha}{n} \left[ {\ln p^n(x)+\Gamma[0,p^n(x)]+\gamma_{\rm E}} \right] \, ,
\quad \mbox{where} \quad 
 \Gamma[0,z]=\int_z^{+\infty} \frac{e^{-z'}}{z'} \, d z'\, , 
\ee
and $\gamma_{\rm E}$ is the Euler-Mascheroni constant. 
Depending on the angular sector, $H(z)$ in \eqref{Hx} hosts a pole or an essential singularity.

We first 
divide the complex plane in regions $\omega^{\epsilon}_{j}$ defined as follows, 
\begin{equation}\label{unstable}
\omega^{\epsilon}_{j}:(2j-3)\frac{\pi}{4n}+\frac{\epsilon}{2}<\theta<(2j-1)\frac{\pi}{4n}-\frac{\epsilon}{2},\quad j=0,\,\cdots,\,4n-1 \, .
\end{equation}
It is known 
that in 
 the regions with $z\in\omega^{\epsilon}_{j\text{ even}}$, an asymptotically polynomial (of degree $n$) analytic form factor behaves asymptotically worse than \cite{Kuzmin}
\begin{equation}
\exp[\exp(A|z|^\rho)], \,\, A>0,  \,\, \rho>0 \, .
\end{equation}
Hence, they host an essential singularity for $|z|\rightarrow\infty$,
while regions with $z \in \omega^{\epsilon}_{j\text{ odd}}$ are s.t. \cite{Tomboulis:1997gg}
\begin{equation}
|f(z)|\stackrel{|z|\rightarrow\infty}{\longrightarrow}0 \, ,
\end{equation}
and include subregions where the form factor behaves as a polynomial.

\subsection{Asymptotically exponential form factors}
Another class of form factors introduced in nonlocal gravity by Krasnikov in $1987$ \cite{Krasnikov:1987yj} is:
\be
H(\Box)= e^{(-\Box)^n} \, , \quad n \in \mathbb{N} \, .
 \label{kra}
\ee
Unfortunately the above form factors are not suitable for gravitational and gauge theories because incompatible with the power counting renormalizability theorems. However, asymptotically exponential form factors are sufficient to make finite, for example, scalar field theories.

\section{Stability problems around a conformally flat background} \label{stab_problems}
The EoMs \eqref{dSEOM} for the perturbations are obtained performing the second order variation of the action 
\eqref{NLtheory} with form factors \eqref{form_factor_S}, \eqref{renormal_form_factor}. 
 The outcome of the expansion around a conformally flat background, which includes the inflationary quasi-dS case, reads \cite{Koshelev:2016xqb, SravanKumar:2018dlo}: 
\begin{align}\label{variations}
    \delta^{2}S_{0}&=-\frac{1}{2}\int d^{4}x\sqrt{-\bar{g}}\,\tilde\phi\left\{\left(\bar{\square}+\frac{\Bar{R}}{3}\right)\left[2\gamma_{1} (\Bar{R}+3M^{2})-2(3\Bar{\square}+R)\gamma_{\rm S}\left(\frac{\bar{\square}}{\Lambda_{*}^{2}}\right)\right]\right\}\tilde\phi \, ,  \\
    \delta^{2}S_{\perp}&=\frac{1}{4}\int d^{4}x\sqrt{-\bar{g}}\, h_{\mu\nu}^{\perp}\Bigg\{\left(\bar{\square}-\frac{\Bar{R}}{6}\right)\times\nonumber \\
    &\hspace{3.4cm}\left[\gamma_{1} (\Bar{R}+3M^{2})+\left(\Bar{\square}-\frac{R}{3}\right)\gamma_{\rm C}\left(\frac{\bar{\square}}{\Lambda_{*}^{2}}+\frac{\bar{R}}{3\Lambda_{*}^{2}}\right)\right]\Bigg\}h^{\perp\mu\nu} \, ,
\end{align}
where $\tilde\phi=\sqrt{{3}/{32}} \, \phi$ is the canonically normalized scalar and $h^{\perp}_{\mu\nu}$ the transverse and traceless part of the metric fluctuation $h_{\mu\nu}=g_{\mu\nu}-\bar g_{\mu\nu}$. 
In the limit $\bar R\gg 6M^2$ and with $\gamma_{\rm C}$ defined in (\ref{renormal_form_factor}), the EoMs for the gravitational perturbations read: 
\begin{align}
\label{dSEOM_3}
    \frac{1}{2}\left(\bar{\square}-\frac{\bar R}{6}\right)\left[e^{H_2\left(\bar{\square}-\frac{\bar R}{3}\right)}+\frac{\bar R}{3M^2}\right]h_{\mu\nu}^{\perp}=0 \, ,
    \end{align}
where $H_2(\Box)$ is defined in \eqref{H2}.
For future reference we define:
\be
f(\bar{\square}) \equiv  \frac{1}{2}\left(\bar{\square}-\frac{\bar R}{6}\right)\left[e^{H_2\left(\bar{\square}-\frac{\bar R}{3}\right)}+\frac{\bar R}{3M^2}\right] \, . 
\label{effeStro}
\ee

\begin{figure}
\label{bound_exp_tomb}
 \centering
    \includegraphics[width=0.479\textwidth]{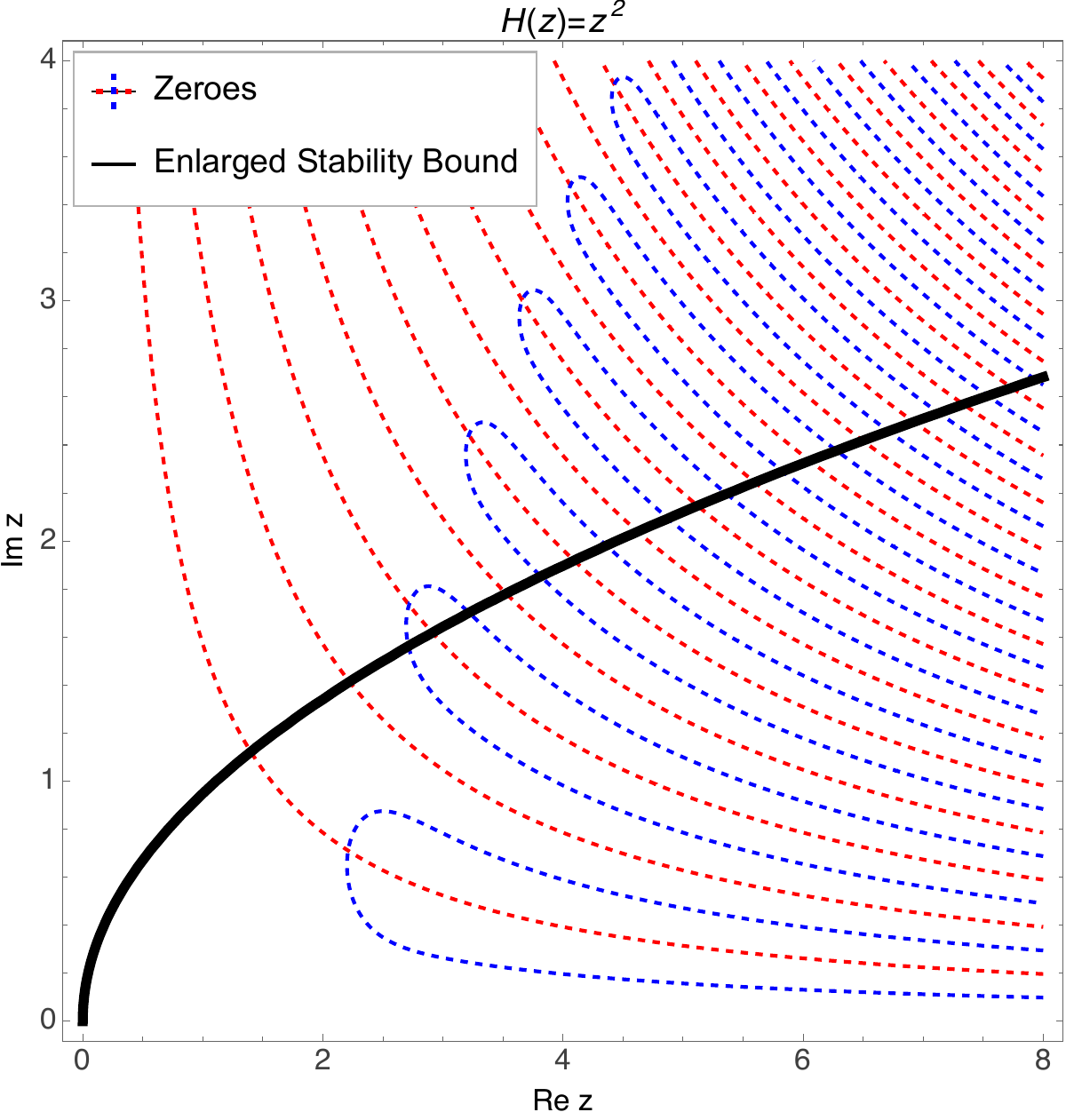}
    \hfill
    \includegraphics[width=0.49\textwidth]{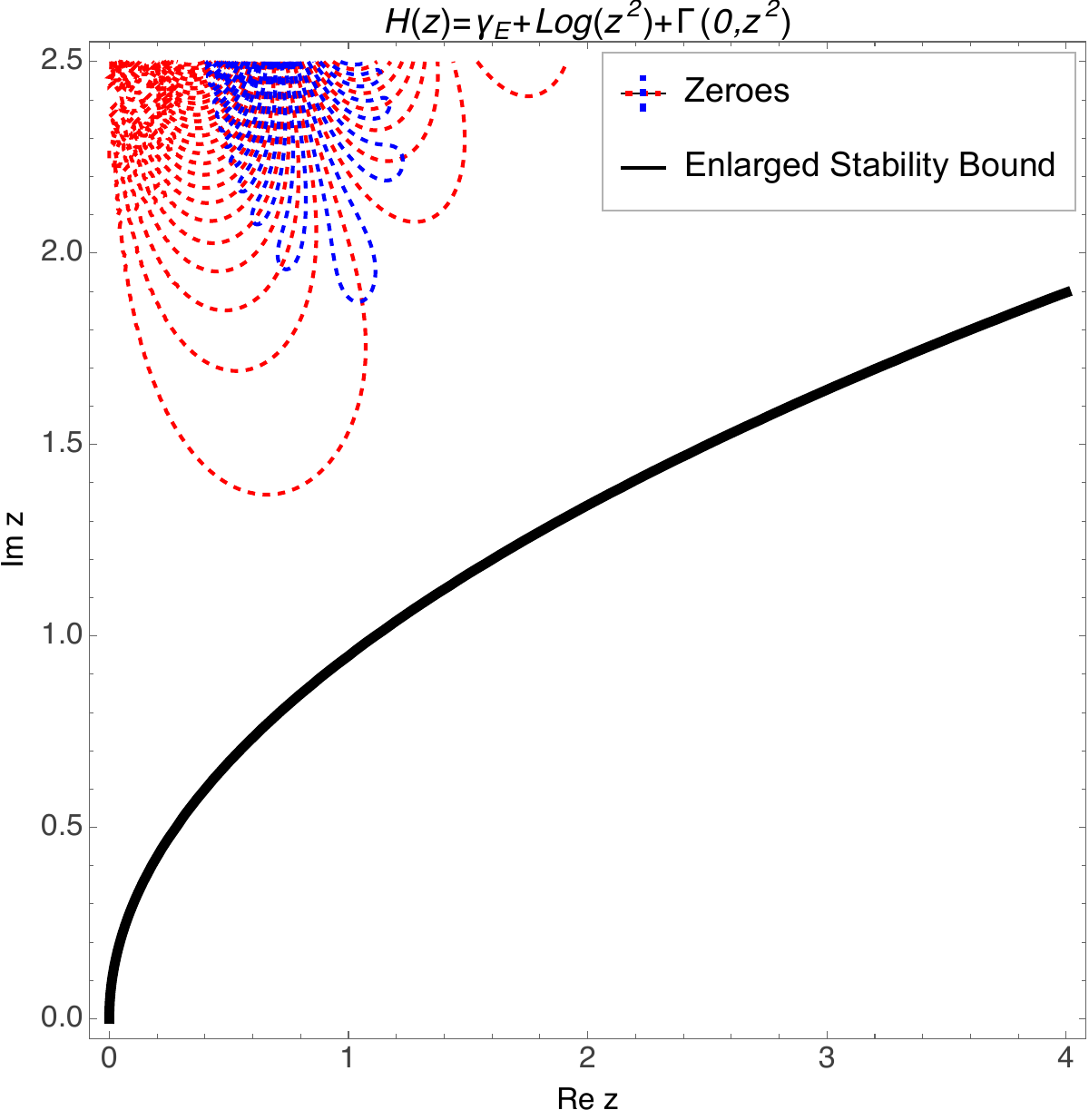}
    \caption{\label{bound_z2} 
    The intersections between the dotted blue curves and dotted red curves, respectively denoting the real and imaginary part of \eqref{stab_eq}, provide the zeros $z_i$. The form factors of interest are written in terms of $z\equiv z_i$ in \eqref{perturb_EOM_simpl}: on 
    the \textbf{left} the form factor is $\exp(z^2)$, on the \textbf{right} is \eqref{H2} with $\alpha=1,\,n=2$. Zeroes outside the enlarged stability bound \eqref{ESB}, namely the parabola delimited in black, are unstable.}
\end{figure}

Following \cite{Koshelev:2020fok}, 
we notice that the function $f(z)$ ($z\equiv \bar{\square}$) in (\ref{effeStro}) is an entire function. Moreover, we  approximate the quasi-dS spacetime with the de Sitter spacetime, namely from now on we work on a \emph{maximally symmetric spacetime} with $\bar R = {\rm constant}$.
Therefore, we implement the following procedure. 
    \begin{enumerate}
    \item We expand $h^\perp_{\mu\nu}$ in
   \eqref{dSEOM_3} on a basis of eigenfunctions $h_i$ of the d'Alembertian operator in de Sitter spacetime  
   with eigenvalues $m_i^2 \in \mathds{R}$.  
\item   The d'Alembertian operator in Minkowski spacetime is diagonal with real eigenvalues $k_\mu$ respect to the Fourier Hilbert's space, namely the basis of plane waves coincides with the spectrum of eigenstates of the operator. 
Using the same basis, we can look for solutions of the following equations,
\be
(\Box^2 +M^4) \Phi = 0 \, ,
\ee
which gives rise to an algebraic equation in the Fourier space, i.e.  
\be 
k^4 + M^4 = 0 ,
\ee
whose solutions are four complex conjugate roots.
Therefore, the solutions relative to such complex eigenvalues come out the Fourier's space because are not normalizable.

The very same procedure applies to the d'Alembertian operator in de Sitter spacetime. The Hilbert space is known and the spectrum is real. Hence, we will proceed by expanding the equation (\ref{dSEOM_3}) respect to such basis and, afterwards, analytically continuing into the complex plane the corresponding algebraic equation. 
     Hence, expanding $h_{\mu\nu}$ in eigenfunctions of the d'Alembertian operator in de Sitter spacetime as follows, 
   \be
          \bar\square h_i=m_i^2 h_i, \quad m_i^2\in\mathds{R} 
          \quad \Longrightarrow \quad 
          h^{\perp}_{\mu\nu}=\sum_i c_{\mu\nu}^ih_i 
        \, , 
   \ee
    the differential equation (\ref{dSEOM_3}) turns into:
    \be
          f(\bar{\square})h^{\perp}_{\mu\nu} = \sum_ic_{\mu\nu}^if(m_i^2)h_i=0 \,\, \implies  \,\, f(m_i^2)=0 \, .
    \ee
    Assuming $m_i^2\in\mathds{C}$, 
    we will show that the algebraic equation $f(m_i^2)=0$  has an infinite number of complex conjugate roots spread over the entire complex plane. 
   
    \item We implement the de Sitter stability condition derived in \cite{Koshelev:2020fok}, which binds the imaginary and real parts of the eigenvalues $m_i^2$ (a representation of the stability region, in terms of $z_i=\frac{m_i^2-\bar R/3}{\Lambda_{*}^2}$, is given in Fig.\ref{bound_z2}), i.e. 
  
\begin{equation}\label{Kosce}
   [\Im(m_i^2)]^2 < 9H^2\Re(m_i^2) \, .
\end{equation} 

    \end{enumerate}
According to the last statement of 2$.$, which we are going to prove shortly, there are always mass perturbations outside the stability region (\ref{Kosce}). Hence, the background metric is unstable. 
    One can think the instability to be welcome to get out the de Sitter stage. However, since $f(z)=0$ has an infinite number of roots up to an arbitrary large value, the lifetime of the de Sitter spacetime is identically zero.
    Therefore,  the theory of Section \ref{stability} can not drive the Starobinsky inflation. 
 
 Regarding the result 2., we recall that the algebraic equation under investigation is:
\begin{align}\label{weier}
    f(m_i^2)=\left(m_i^2-\frac{\bar R}{6}\right) \left[ e^{H_{2}\big(m_i^2-\frac{\bar R}{3}\big)}+4\frac{H_{0}^{2}}{M^{2}}\right] =0 \, .
\end{align}
 
    Besides the solutions due to the standard dS kinetic operator $\bar{\square}-\bar R/6$, 
    we have a solution to \eqref{weier} for any root $z_i$ of the following equation, 
    \begin{equation}
    \label{perturb_EOM_simpl}
    e^{H_2(z_i)}+C^2=k,\quad z_i=\frac{m_i^2-\bar R/3}{\Lambda_{*}^2}, \quad C^2=4\frac{H_{0}^{2}}{M^{2}}, \quad k=0\, ,
\end{equation}
where we introduced the redundant parameter $k$ to facilitate the following proof.

Now we can show that there exist infinitely many solutions of \eqref{perturb_EOM_simpl}, 
provided $H_2(z)$ to be an entire function, invoking the \emph{Great Picard's Theorem}:
If an analytic function $f(z)$ has an essential singularity at a point $w$, then, on \underline{any} punctured \footnote{Namely $w$ itself is not included.} neighborhood of $w$, the function $f(z)$ takes on all possible complex values, with at most a single exception, \underline{infinitely often}. 

The idea is the following: thanks to \cite{Tomboulis:1997gg} we know where to find essential singularities $w$ for an asymptotic polynomial $\exp H_2$ \eqref{H2}. We also know that $e^{H_2(z)}\neq0 \quad\forall z$ because $H_2$ is entire. Hence, taken a neighborhood $I(w)$ around $w$, we know from the previous theorem that 
$f[I(w)]$ has image $\mathds{C}\setminus \{k=C^2\}$, where any value in the image of the function is taken infinitely many times, including our case of interest $f(z^*)=0$.

Hence, regarding our case, we are stating that 
\begin{equation}\label{stab_eq}
e^{H_2(z)}+C^2=0, \quad C^2=\dfrac{240}{3}
\end{equation}
has infinitely many solutions, provided the hypotheses of the theorem are satisfied. Those solutions are located in the neighborhoods of essential singularities, which are points at infinity approached from given angular directions. In the asymptotic exponential cases \eqref{Hx} they appear in \eqref{unstable} with $j$ odd.

Finally, since we are free to choose the neighborhood as we prefer, as long as it is extended upon $|z|=\infty$ on the correct angular directions, it is straightforward to avoid the stability region choosing a suitable lower bound for the modulus. 

We remark that generally we are not able to say exactly whose $z_i$ (respectively $m_i^2$) satisfy \eqref{perturb_EOM_simpl} (respectively \eqref{weier}). 
Conversely, we are sure of the existence of regions where there are infinitely many of these solutions.
Even worse, from the previous argument there will be no upper limit on the masses of unstable perturbations, from which follows that de Sitter lifetime is exactly zero.

Similar considerations can be done for exponential form factors.
To get a taste of the previous abstract analysis, we show in Figure \ref{bound_z2} a subset of the solutions of \eqref{stab_eq} for a purely exponential form factor $\exp(z^2)$ and for an asymptotic polynomial of the form \eqref{H2} with $\alpha=1,\,n=2$. In order to improve the visibility of the stability region, we defined an \emph{enlarged stability bound} such that
\begin{equation}\label{ESB}
    \Im(z_i)^2<0.9\,\Re\left(z_i+\dfrac{\bar R}{3M_{\rm p}^2}\right)
\end{equation}
for $z_i$ in \eqref{perturb_EOM_simpl} with $\Lambda_\star=M_{\rm p}$.
A solution outside the enlarged stability bound will necessarily be outside the stability region \eqref{Kosce}: recalling that $H^{2}=\bar R/12$ with $\bar R\simeq240 M^2;\,M=1.3\cdot 10^{-5}
M_{\rm p}$ it follows

\begin{equation}\label{ESB2}
  0.9\,  \Re\left(z_i+\dfrac{\bar R}{3M_{\rm p}^2}\right)>\dfrac{9H^2}{M_{\rm p}^2}\,\Re\left(z_i+\dfrac{\bar R}{3M_{\rm p}^2}\right) .
\end{equation}

Notice that the previous considerations would have dramatically changed if the form factor were a polynomial, i.e. 
\begin{equation}
p(z)=\sum_{i=1}^{n}a_i z^i \, , 
\end{equation}
in place of $\exp {H_2(z)}$. Indeed, it has a finite number of poles and, with a fine-tuning the coefficients $\{a_i\}$, we can push every pole inside the parabola-shaped region \eqref{Kosce}, ensuring the presence of only stable fluctuations.

The previous observations can also be stated as follows: any transcendental entire function 
$\exp H_2(z)$, which is asymptotically exponential (\ref{kra}) 
or asymptotically polynomial \eqref{Hpoly}, violates \eqref{Kosce} for an infinite number of complex masses, provided that the perturbations obey an equation of the form \eqref{dSEOM_3}. 

To date we are not aware of form factors at the same time asymptotically polynomial and with essential singularities only in the stability region.

\section{Non-renormalizability of nonlocal $f(R)$ in the Jordan frame}\label{JF_details}
We here show that the nonlocal extension \eqref{action} of $\mathcal{L}_{\text{loc}} = M_{\rm p}^2 f(R)/2$ 
is non-renormalizable and, thus, we can not achieve the ultraviolet completion of $f(R)$ in the Jordan frame. 

According to the recipe \eqref{action}, the nonlocal extension takes the form:
\begin{align}
     S&=\int d^4x \sqrt{-g} \left[\dfrac{M_\mathrm{p}^2}{2}f(R) + \frac{2}{M_{\rm p}^2} 
    E_{\mu\nu}
    F^{\mu\nu\rho\sigma}(\Delta)E_{\rho\sigma}\right], \nonumber \\
     E_{\mu\nu}&= \frac{M_\mathrm{p}^2}{2} \left[ f'(R) R_{\mu\nu}-\frac{1}{2}f(R)g_{\mu\nu}-(\nabla_{\mu}\nabla_{\nu}-g_{\mu\nu}\Box)f'(R) \right]  \, ,
\end{align}
where $f'(R) \equiv df/dR$, and the local theory is defined by the analytic function:
 \be
 f(R)=R+\epsilon R^2+\sum_{n=3}^{N}a_n R^n \, . 
 \ee

The kinetic term of the metric perturbation around Minkowski comes from operators at most made of two curvature tensors, 
\be
&&    S =  \frac{M_{\rm p}^2}{2} \!\! 
\int d^4x \sqrt{-g}\Big[R+\epsilon R^2+ \Big(R_{\mu\nu} + 2 \epsilon R R_{\mu\nu} 
- \dfrac{1}{2} \left(R + \epsilon R^2\right) g_{\mu\nu}  \nonumber \\
&& \hspace{0.5cm}
\quad- (\nabla_\mu \nabla_\nu-g_{\mu\nu}\Box) \left(2\epsilon R + 3 a_3 R^2\right) + O(\mathcal{R}^3 )\Big) 
F^{\mu\nu\rho\sigma}(\Delta)\times
\nonumber \\
&&
\hspace{0.75cm}
\Big(R_{\rho\sigma} + 2 \epsilon R R_{\rho \sigma}
- \dfrac{1}{2}\left( R + \epsilon R^2 \right)  g_{\rho\sigma}-(\nabla_\rho \nabla_\sigma - g_{\rho\sigma}\Box)
\left(2\epsilon R + 3 a_3 R^2\right) + O\left(\mathcal{R}^3\right)\Big) 
\Big] \nonumber \\
  &&  \hspace{0.35cm}
  = \int d^4x \sqrt{-g}\Big[R+R\gamma_0(\Box)R+R_{\mu\nu}\gamma_2(\Box)R^{\mu\nu}+O\left(\mathcal{R}^3\right)\Big],\label{SZoppa}
\ee
where in the last equality we have replaced $\Delta$ with $\Box$ because we are mainly interested in the propagator, while $\gamma_0(\Box)$ and $\gamma_2(\Box)$ are uniquely fixed by the tree-level unitarity, namely the spectrum consists of at most the graviton and the scalar curvaton \cite{Briscese:2012ys,Briscese:2013lna}, and requiring the same scaling of the two components of the propagator at high energy. The result in $D=4$ reads \cite{Modesto:2016max}:
\be
&& \gamma_2 = \frac{e^{H_2} - 1}{\Box} \, , \nonumber \\
&& \gamma_0 = -  \frac{ 2(e^{H_0} - 1) + 4( e^{H_2}  - 1)}{12 \Box} \, , 
\label{gamma02D4}
\ee
where the pole of the curvaton is implemented by the following replacement,
\be
e^{H_0} \quad \rightarrow \quad e^{\tilde{H}_0} ( 1 + \alpha \epsilon \Box) \, . 
\label{curvatonic}
\ee
Given the above definitions of $\gamma_0$ and $\gamma_2$ the propagator is (\ref{appendix_propagator}) for $\alpha = - 6$. In order to have the same ultraviolet scaling for the spin two and spin zero components of the propagator, the asymptotic polynomial limit of $\exp \tilde{H}_0$ must have one less degree respect to $\exp {H}_2$. 
Now we have to determine $F$ consistently with the last identity in (\ref{SZoppa}) and the form factors (\ref{gamma02D4}). 
We decompose the form factor $F$ as follows,
\be
F^{\mu\nu\rho\sigma}(\Box) = g^{\mu\nu}g^{\rho\sigma}f_1(\Box)+g^{\mu\rho}g^{\nu\sigma}f_2 (\Box) \, ,
\label{Fciccio}
\ee
and we determine $f_1$, $f_2$ in terms of $\gamma_0$, $\gamma_2$ by comparing the last and the second to last lines in (\ref{SZoppa}). 
After a long and tedious expansion we get:
\be
f_2 = \gamma_2(\Box) \, , \qquad f_1 = \frac{\gamma_0(\Box) - \epsilon - 4 \epsilon \Box \gamma_2(\Box) ( 3 \epsilon \Box -1 )}{12 \epsilon \Box (3 \epsilon \Box -1) + 1}.
\ee

Replacing (\ref{gamma02D4}) in $f_1$ we get:
\be
f_1 = \frac{- e^{\tilde{H}_0} ( 1 + \alpha \epsilon \Box ) +1 - 6 \epsilon \Box }{6 \Box [ 12 \epsilon \Box (3 \epsilon \Box -1) + 1 ] } 
- \frac{e^{H_2} - 1}{3 \Box} 
\, .
\ee
Now in order to ensure analyticity of the form factor $f_1$, we make the choice $\alpha = -6$. Therefore, according to (\ref{localR^2}), $\epsilon = 1/6 M^2$ and $f_1$ simplifies to:
\be
f_1 = -  \frac{e^{\tilde{H}_0} -1 }{6 \Box \left( 1 - \frac{\Box}{M^2} \right) } 
- \frac{e^{H_2} - 1}{3 \Box} 
\, .
\ee
The analyticity requires the argument of $\tilde{H}_0$ to have a zero at the curvaton pole:
\be
\Box^{n - 1} \left( 1 - \frac{\Box}{M^2} \right)
\ee
Therefore, if we assume $\exp H_2$ to have an asymptotic polynomial scaling of degree $n+1$, we get: 
\be
 \gamma_2 \,\, \rightarrow \,\, \Box^n \, ,  \quad 
 \gamma_0 \,\, \rightarrow \,\, \Box^n \, ,  \quad
  f_1 \,\, \rightarrow \,\, \Box^n \, ,  \quad 
 f_2 \,\, \rightarrow \,\, \Box^n \, .
\ee
Here we are faced with a serious issue. Indeed, while the falloff of the propagator is $1/\Box^{n+2}$ (\ref{propagator}), already in Starobinsky theory we have problematic vertices at least cubic in the curvature, such as $\mathcal{R}f_i(\Box)\mathcal{R}^2$, that scale at least as $\Box^{n+3}$. Therefore, the theory is not renormalizable.   
In general, according to our recipe, the only $f(R)$ theory whose nonlocal uplift has the same maximal number of derivatives in the kinetic term as well as in the vertices is the Einstein-Hilbert action.

\section{Dimensional regularization and cut-off scheme}\label{cutoff}

We here motivate why power law divergences emerging in the cut-off regularization scheme, besides $\log$-divergences, do not affect the classical action.
To be precise, the corresponding counterterms do not come together with a form factor that can modify the equations of motion. 
Since the equations of motion are the same up to an unphysical redefinition of some parameters, then, the power law divergences are immaterial. 
Let us consider a scalar toy model with a derivative vertex, i.e. 
\be
\mathcal{L}_{\rm toy} = \frac{1}{2} \phi (\Box + m^2 )\phi +
\frac{\lambda}{3!} \phi^2 \frac{\Box}{M^2} \phi \, , 
\ee
where $M$ is in the Lagrangian in order to preserve the right mass dimensions, and $[\phi] =  [ \lambda] = 1$.
At one loop we get a quadratic and a quartic divergence in the cut-off according to the number of derivatives acting on the two external legs. The quadratic divergence is, up to a constant coefficient,
\be
\delta^4( p - q ) \, \frac{\lambda^2}{M^4}  \phi^2 p^2 \int^\Lambda d^4 k \frac{k^2}{(k^2+m^2)[ (p-k)^2 + m^2]} 
\quad \stackrel{\rm UV}{\rightarrow}  \quad \delta^4( p - q ) \,  \frac{\lambda^2}{M^4}  \phi^2 p^2 \, \Lambda^2 
\ee
that requires the counterterm:
\be
 \frac{\lambda^2}{M^4} \Lambda^2 \int d^4 x \phi \Box \phi \, ,
\ee
which is just a constant redefinition of the kinetic term, but does not come together with a form factor, namely the kinetic term is exactly the classical one. 

We also get a quartic divergence when no derivatives act on external legs, 
\be
\delta^4( p - q ) \, \frac{\lambda^2}{M^4}  \phi^2  \int^\Lambda d^4 k \frac{k^4}{(k^2+m^2)[ (k+p)^2 + m^2]} 
\quad \stackrel{\rm UV}{\rightarrow}  \quad \delta^4( p - q ) \,  \frac{\lambda^2}{M^4}  \phi^2  \, \Lambda^4 \, , 
\ee
and now the required  counterterm is:
\be
 \frac{\lambda^2}{M^4} \Lambda^4 \int d^4 x \phi  \phi \, .
\ee
The latter is just a redefinition of the mass term, but again it does not produce a form factor that can affect the classical equations of motion.

Therefore, from the quadratic and quartic divergences does not arise any running or new physics, technically there is no form factor or modification of the effective EoM. In short, the quadratic and quartic divergences are harmless. 

\section{Unified nonlocal theory of gravity and matter}\label{ADetails}
We here explicitly provide the variations $E_i$ and the Hessian $\hat{\Delta}_{ij}$ that appear in the nonlocal Lagrangian \eqref{NLGM_fR} and its UV limit \eqref{UV_action}, 
\begin{align}\label{EOM_Hessian0}
\begin{split}
     E_{\varphi}\equiv&\frac{\delta S_{\rm loc}}{\delta\varphi}=\square\varphi-V'(\varphi),\\
     E^{\mu\nu}\equiv&\frac{\delta S_{\rm loc}}{\delta g_{\mu\nu}}=\frac{M_{p}^{2}}{2}\Big(-R^{\mu\nu}+\frac{1}{2}g^{\mu\nu}R\Big)+\frac{1}{2}(\nabla^{\mu}\varphi)(\nabla^{\nu}\varphi)-\frac{1}{4}g^{\mu\nu}(\nabla\varphi)^{2}-\frac{1}{2}g^{\mu\nu}V(\varphi),\\
     \hat{\Delta}_{11}^{\mu\nu,\alpha\beta} =&
\frac{1}{\kappa^2} \Big[ \frac{1}{2} \delta^{\mu\nu,\alpha\beta} \Box
- \frac12 g^{\mu\nu} g^{\alpha\beta} \Box 
+ \frac12 ( g^{\mu\nu} \nabla^{\alpha}\nabla^{\beta}
+ g^{\alpha\beta} \nabla^{\mu}\nabla^{\nu} )\\
&
- \frac14 (
g^{\mu\alpha} \nabla^{\nu}\nabla^{\beta} +  g^{\nu\alpha} \nabla^{\mu}\nabla^{\beta}
+ g^{\mu\beta} \nabla^{\nu}\nabla^{\alpha} +  g^{\nu\beta} \nabla^{\mu}\nabla^{\alpha})
\Big] + \Pi^{\mu\nu,\alpha\beta},\\
\hat{\Delta}_{12}^{\mu\nu} =&  2 \mathcal{G}^{\mu\nu,\lambda\tau}_0 (\nabla_\lambda \varphi) \nabla_\tau - \frac12 g^{\mu\nu}V'(\varphi),\\
\hat{\Delta}_{21}^{\alpha\beta} =&  - 2 \mathcal{G}^{\alpha\beta,\lambda\tau}_0 [(\nabla_\lambda \varphi) \nabla_\tau + (\nabla_\lambda \nabla_\tau \varphi) ] - \frac12 g^{\alpha\beta} V'(\varphi),\\
\hat{\Delta}_{22} =&  \Box - V'' (\varphi),
\end{split}
\end{align}
where we introduced the following definitions, 
\begin{align}
    \begin{split}
\mathcal{G}^{\mu\nu,\alpha\beta}_0 \equiv& \frac12 (\delta^{\mu\nu,\alpha\beta} - \dfrac12 g^{\mu\nu} g^{\alpha\beta} ),\\
\Pi^{\mu\nu,\alpha\beta} \equiv& \frac{1}{\kappa^2} \Big[ \frac14 (R^{\mu\alpha\nu\beta} + R^{\nu\alpha\mu\beta} + R^{\mu\beta\nu\alpha} + R^{\nu\beta\mu\alpha})- \frac12 ( g^{\mu\nu} R^{\alpha\beta} + g^{\alpha\beta} R^{\mu\nu})\\
& + \frac14 \left(g^{\mu\alpha} R^{\nu\beta} + g^{\nu\alpha} R^{\mu\beta} + g^{\mu\beta} R^{\nu\alpha} + g^{\nu\beta} R^{\mu\alpha}   \right)
- \mathcal{G}_0^{\mu\nu,\alpha\beta} R \Big]\\
&+\frac12 \mathcal{G}^{\mu\nu,\alpha\beta}_0 (\nabla \varphi)^2 + \mathcal{G}^{\mu\nu,\alpha\beta}_0 V(\varphi)
+ \frac14 \Big[ g^{\mu\nu} (\nabla^\alpha \varphi) (\nabla^\beta  \varphi) + g^{\alpha\beta} (\nabla^\mu  \varphi) (\nabla^\nu  \varphi) \Big] \\
&- \frac14 \Big[g^{\mu\alpha} (\nabla^\nu  \varphi) (\nabla^\beta  \varphi) + g^{\nu\alpha} (\nabla^\mu  \varphi) (\nabla^\beta  \varphi) + g^{\mu\beta} (\nabla^\nu  \varphi) (\nabla^\alpha \varphi) + g^{\nu\beta} (\nabla^\mu  \varphi) (\nabla^\alpha \varphi)   \Big].
\end{split} 
\end{align}
According to the rescaling of the scalar field as defined before \eqref{NLGM_fRMM}, namely $\varphi \rightarrow M_{\rm p} \varphi$, the EoMs and the Hessian of the local theory change as follows, 
\begin{align}
\label{EOM_Hessian}     
   E_{\varphi}\,\, &\rightarrow \,\,
    \frac{\delta S_{\ell}}{\delta\varphi}= M_{\rm p}^2 \left[ \square\varphi-\mathcal{V'}(\varphi) \right]
    \equiv M_{\rm p}^2 \, E_{\varphi}^\prime 
    , \\
   \label{EOM_grav}
 E^{\mu\nu} \,\, &\rightarrow   \,\, 
 \frac{\delta S_{\rm loc}}{\delta g_{\mu\nu}} = 
 M_{\rm p}^{2} \left[ \frac{1}{2} \Big(-R^{\mu\nu}+\frac{1}{2}g^{\mu\nu}R\Big)+\frac{1}{2}(\nabla^{\mu}\varphi)(\nabla^{\nu}\varphi)-\frac{1}{4}g^{\mu\nu}(\nabla\varphi)^{2}
 -\frac{1}{2}g^{\mu\nu} 
 \mathcal{V}(\varphi) \right] 
 \\&\hspace{1.65cm}\equiv M_{\rm p}^2 \, E^{\prime \mu\nu}
 ,\\    
  \hat{\Delta}_{11}^{\mu\nu,\alpha\beta} \,\, &\rightarrow \,\, 
\frac{M_{\rm p}^2}{2} \Big[ \frac{1}{2} \delta^{\mu\nu,\alpha\beta} \Box
- \frac12 g^{\mu\nu} g^{\alpha\beta} \Box 
+ \frac12 ( g^{\mu\nu} \nabla^{\alpha}\nabla^{\beta}
+ g^{\alpha\beta} \nabla^{\mu}\nabla^{\nu} ) 
\\
& \hspace{0.7cm}- \frac14 (
g^{\mu\alpha} \nabla^{\nu}\nabla^{\beta} +  g^{\nu\alpha} \nabla^{\mu}\nabla^{\beta}
+ g^{\mu\beta} \nabla^{\nu}\nabla^{\alpha} +  g^{\nu\beta} \nabla^{\mu}\nabla^{\alpha})
\Big] + \Pi^{\mu\nu,\alpha\beta}
\equiv M_{\rm p}^2 \hat{\Delta}_{11}^{\prime \mu\nu,\alpha\beta} 
,\\
\hat{\Delta}_{12}^{\mu\nu} \,\, &\rightarrow \,\, M_{\rm p}^2 \left[   2 \mathcal{G}^{\mu\nu,\lambda\tau}_0 (\nabla_\lambda \varphi) \nabla_\tau - \frac12 g^{\mu\nu} \mathcal{V}'(\varphi) \right] 
\equiv M_{\rm p}^2 \hat{\Delta}_{12}^{\prime \mu\nu} 
,\\
\hat{\Delta}_{21}^{\alpha\beta} 
 \,\, &\rightarrow \,\, M_{\rm p}^2 \left[ 
   - 2 \mathcal{G}^{\alpha\beta,\lambda\tau}_0 [(\nabla_\lambda \varphi) \nabla_\tau + (\nabla_\lambda \nabla_\tau \varphi) ] - \frac12 g^{\alpha\beta} \mathcal{V}'(\varphi) \right]
   \equiv M_{\rm p}^2 \hat{\Delta}_{21}^{\prime \alpha\beta} 
   ,\\
\\
\hat{\Delta}_{22}  \,\, &\rightarrow \,\, M_{\rm p}^2 \left[  \Box - \mathcal{V}'' (\varphi) \right] 
\equiv M_{\rm p}^2 \hat{\Delta}_{22}^{\prime} 
\, ,
\end{align}
where the previous definitions turn into:
\begin{align}
    \begin{split}
\mathcal{G}^{\mu\nu,\alpha\beta}_0 \equiv& \frac12 (\delta^{\mu\nu,\alpha\beta} - \dfrac12 g^{\mu\nu} g^{\alpha\beta} ),\\
\Pi^{\mu\nu,\alpha\beta}\,\, \rightarrow  & \,\, M_{\rm p}^2  \Big[ \frac14 (R^{\mu\alpha\nu\beta} + R^{\nu\alpha\mu\beta} + R^{\mu\beta\nu\alpha} + R^{\nu\beta\mu\alpha})- \frac12 ( g^{\mu\nu} R^{\alpha\beta} + g^{\alpha\beta} R^{\mu\nu}\\
& + \frac14 \left(g^{\mu\alpha} R^{\nu\beta} + g^{\nu\alpha} R^{\mu\beta} + g^{\mu\beta} R^{\nu\alpha} + g^{\nu\beta} R^{\mu\alpha}   \right)
- \mathcal{G}_0^{\mu\nu,\alpha\beta} R \Big]\\
&+\frac12 \mathcal{G}^{\mu\nu,\alpha\beta}_0 (\nabla \varphi)^2 + \mathcal{G}^{\mu\nu,\alpha\beta}_0 \mathcal{V}(\varphi)
+ \frac14 \Big[ g^{\mu\nu} (\nabla^\alpha \varphi) (\nabla^\beta  \varphi) + g^{\alpha\beta} (\nabla^\mu  \varphi) (\nabla^\nu  \varphi) \Big] \\
&- \frac14 \Big[g^{\mu\alpha} (\nabla^\nu  \varphi) (\nabla^\beta  \varphi) + g^{\nu\alpha} (\nabla^\mu  \varphi) (\nabla^\beta  \varphi) + g^{\mu\beta} (\nabla^\nu  \varphi) (\nabla^\alpha \varphi) + g^{\nu\beta} (\nabla^\mu  \varphi) (\nabla^\alpha \varphi)   \Big].
\end{split} 
\end{align}
For completeness we write explicitly the higher derivative operator that can give contributions to the divergences in the quantum effective action,
\be
   S_{\rm UV}=  \alpha \int d^4 x\sqrt{-g} \Bigg[
  \sum_{k=0}^{n}E^\prime_{i} \left( \frac{\Delta^{\prime k}}{M_{\rm p}^{2k} } \right)^{ij} \!\!E^\prime_{j}\Bigg]  .
    \label{NLGM_fRMMUV}
\ee

\bibliographystyle{JHEP}
\bibliography{biblio}

\end{document}